\documentclass[%
 superscriptaddress,
preprint,
 amsmath,amssymb,
 aps,longbibliography
]{revtex4-1}

\usepackage{graphicx}
\usepackage{dcolumn}
\usepackage{bm}
\usepackage{hyperref}
\usepackage[mathlines]{lineno}
\usepackage[bottom]{footmisc}
\usepackage{booktabs}
\usepackage[super]{nth}%

\begin{document}

\title{Aperiodic photonics of elliptic curves}
\author{Luca  Dal Negro}
\email{dalnegro@bu.edu}
\affiliation{Department of Electrical and Computer Engineering, Boston University, Boston, Massachusetts, 02215, USA.}
\affiliation{Division of Material Science and Engineering, Boston University, Boston, Massachusetts, 02215, USA.}
\affiliation{Department of Physics, Boston University, Boston, Massachusetts, 02215, USA}
\author{Yuyao Chen}
\affiliation{Department of Electrical and Computer Engineering, Boston University, Boston, Massachusetts, 02215, USA.}
\author{Fabrizio Sgrignuoli}
\affiliation{Department of Electrical and Computer Engineering, Boston University, Boston, Massachusetts, 02215, USA.}
%

\begin{abstract}
In this paper we propose a novel approach to aperiodic order in optical science and technology that leverages the intrinsic structural complexity of certain non-polynomial (hard) problems in number theory and cryptography for the engineering of optical media with novel transport and wave localization properties. In particular, we address structure-property relationships in a large number (900) of light scattering systems that physically manifest the distinctive aperiodic order of elliptic curves and the associated discrete logarithm problem over finite fields. Besides defining an extremely rich subject with profound connections to diverse mathematical areas, elliptic curves offer unprecedented opportunities to engineer light scattering phenomena in aperiodic environments beyond the limitations of traditional random media. Our theoretical analysis combines the interdisciplinary methods of point patterns spatial statistics with the rigorous Green's matrix solution of the multiple wave scattering problem for electric and magnetic dipoles and provides access to the spectral and light scattering properties of novel deterministic aperiodic structures with enhanced light-matter coupling for nanophotonics and metamaterials applications to imaging and spectroscopy.
\end{abstract}

\maketitle
\section{Introduction}

Stimulated by P. W. Anderson's realization that strong disorder can inhibit electronic transport \cite{Anderson}, the study of quantum and classical waves in disordered media with randomly fluctuating potentials has unveiled profound analogies between the electronic and the optical behavior of complex materials \cite{DiederikPhotonic,LagendijkToday,Sheng,BertolottiNecklace}. Moreover, understanding wave transport and localization phenomena in aperiodic optical media provides opportunities to tailor their optical density of states and to enhance light-matter interactions for the engineering of novel active photonic devices. Specifically, the study of multiple light scattering in random media led to the demonstration of random lasers with both uniform \cite{Cao,CaoReview,Leonetti,Lawandy1,Lawandy2} and correlated disorder \cite{ChenScirep} , as well as to remarkable advances in optical imaging \cite{BertolottiNature,Mosk,SebbahImaging,Katz,Vellekoop} and spectroscopy \cite{Redding,Redding2}. However, despite a sustained research effort Anderson localization of optical waves remains an elusive phenomenon since it does not occur in open-scattering random media when the vector nature of light is taken into account. 

The lack of Anderson localization in optical random media is attributed to the detrimental effects of near-field coupling of electromagnetic waves confined at the sub-wavelength scale between scatterers in dense systems \cite{SkipetrovNJP,SkipetrovPRL,Bellando}. Due to the uncorrelated nature of the disorder, it is difficult to overcome this problem and to establish simple design rules for the optimization of uniform random media, often limiting their applications to optical device engineering. As a result, there is currently a compelling need to develop optical media that are deterministic in nature while at the same time sufficiently structurally-complex to offer an alternative route to achieve stronger light localization effects compared to uniform random systems. In response to these challenges, deterministic aperiodic structures have been developed. Deterministic structures with aperiodic though long-range ordered distributions of scattering potentials have a long history in the electronics and optics communities due to significant advantages in design and compatibility with standard fabrication technologies compared to random systems \cite{MaciaBook,LucaBook}. These structures manifest unique spectral characteristics that lead to physical properties that cannot be found in either periodic or uniform random media, such as multifractal density of eigenstates with varying degrees of spatial localization, known as critical modes \cite{DalNegro_PRLFibo,DalNegro_Crystals,DalNegro_LaserPhoton,MaciaCritical,Ryu}, anomalous photon transport regimes \cite{DalNegroScirep,SokolovFractional}, and distinctive wave localization transitions \cite{SgrignuoliVogel,FroufePNAS}. 
Critical modes feature highly-fragmented envelopes characterized by local power-law scaling that reflects the multi-scale geometry of certain deterministic aperiodic potentials that found recent applications to light emission and lasing, optical sensing, photo-detection, and nonlinear optical devices \cite{DalNegro_PRLFibo,Gellermann,Vardeny,Razi,DalNegro_LaserPhoton,Trevino,Lifshitz,Shalaev,Mahler,Capretti,Pecora}. 
An alternative strategy relies on the engineering of scattering structures based on the distinctive aperiodic order, unpredictability, and complexity that naturally arise in the context of number theory \cite{Schroeder,LucaBook,Wang}.  This profound and deeply-fascinating field of mathematics provides paradigmatic examples of the
subtle interplay between structure and randomness \cite{Miller,Schwarz}.
Examples include the aperiodic distribution of prime numbers and their algebraic field generalizations, 
the almost-periodicity characteristic of arithmetic functions,
aperiodic primitive roots and quadratic residue sequences, 
the intricate behavior of Dirichlet $L$-functions (which include the Riemann's $\zeta$ function), and the distribution of binary digits in Galois fields, just to name a few.

In this paper we introduce a novel class of deterministic structures that manifest the distinctive aperiodic order of elliptic curves and the associated discrete logarithm problem over finite fields. In particular, using the Green's matrix formalism we systematically study the spectral and localization properties of their scattering resonances with respect to uniform random systems and we address distinctive structure-property relationships using the methods of point patterns spatial analysis. Finally, we present an extension of the coupled electric dipole method that includes the scattering contribution of magnetic modes, which are important for the accurate design of aperiodic arrays with finite-size dielectric nanoparticles. Specifically, we apply our method to the study of the scattering spectra and the forward/backward scattering response of elliptic curves and discrete logarithm structures composed of $TiO_{2}$ nanoparticles of sub-wavelength dimensions. 

Our results demonstrate that the light scattering properties of particle arrays designed according to the proposed elliptic curve approach are distinctively different from the ones of uniform random systems, despite close similarities are observed in both point pattern and spectral statistics. Based on the analysis of 900 different structures, we show that at small values of optical density the distributions of the level spacing of the complex resonances of elliptic curve structures (and of their discrete logarithm) is described by critical statistics, differently from the usual case of diffusive transport regime encountered in uniform random systems. Random systems exhibit critical statistics only at the density corresponding to the localization threshold, where all the eigenmodes are known to exhibit fractal scaling \cite{Zharekeshev}. In contrast, here we show numerically that elliptic curve structures display critical spectral statistics over a large range of densities until they transition into a more localized transport regime described by Poisson statistics at very large densities. Our comprehensive analysis also indicates that elliptic curve structures feature a much smaller fraction of sub-radiant proximity resonances compared to traditional random systems, resulting in significantly increased modal lifetimes and enhanced light-matter coupling. Finally, we design elliptic curve and discrete logarithm arrays of $TiO_{2}$ nanoparticles with resonant scattering across the visible spectrum and demonstrate a large tunability of the spectral width of their  back-scattered radiation. 

Our findings not only underline the importance
of structural correlations in elliptic curve-based structures for the improvement of photonic systems but also show that the solution of the associated wave scattering problem reveals remarkable differences in the scattering and localization properties that may become important for the optical identification of vulnerabilities in elliptic-curve cryptosystems. 

\section{Elliptic curves and  discrete logarithm structures}\label{ECDL_implementation}

\begin{figure}[ht]
\centering
\includegraphics[width=12cm]{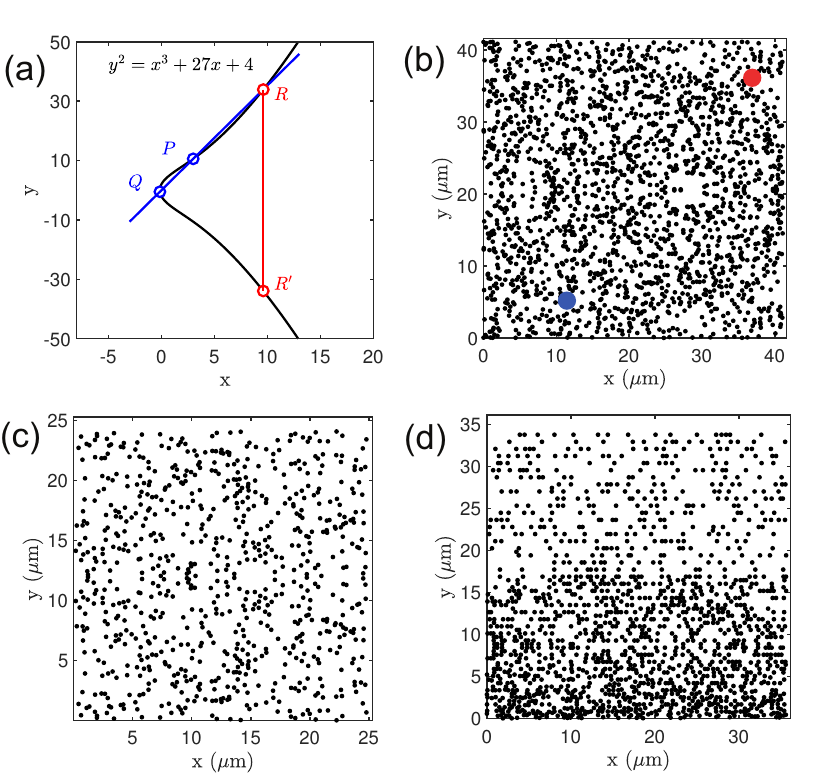}
\caption{(a) Continuous elliptic curve generated by Eq.(\ref{eq1}) when the coefficients $A$ and $B$ are equal to 27 and 4, respectively. The sum operation on elliptic curve $R^{\prime}=P\oplus Q$ is also shown. (b) Point pattern generated from the continuous curve of panel (a) defined over the finite field $\mathbb{F}_{2111}$ rescaled to have an average interparticles separation equal to 450$nm$. The red and blue point marker identifies two representative points $W$ and $M$, respectively. Panels (c) and (d) show the point patterns generated by solving the discrete log problem $W=kM$. Specifically, panels (c) and (d) are characterized by the coordinates $(M_x; k)$ and $(M_y; k)$, respectively. Also for these geometries we rescaled the generated point patterns to have an average interparticles separation equal to 450$nm$.}
\label{Fig1}
\end{figure}

An elliptic curve $E(\mathbb{K})$ over a number field $\mathbb{K}$ is a non-singular curve (i.e. with a unique tangent at every point) with points in $\mathbb{K}$ that are the solutions of a cubic equation. Therefore, elliptic curves can be thought of as the set of solutions in the field $\mathbb{K}$ of equations in the form \cite{Hoffstein}:
\begin{equation}\label{eq1}
y^2=x^3+Ax+B
\end{equation}
where the coefficients $A$ and $B$ belong to $\mathbb{K}$ and satisfy the non-singular condition $\Delta_{E}=4A^3+27B^2\neq{0}$ for the discriminant $\Delta_{E}$ that excludes cusps or self-intersections (i.e., knots) \cite{Hoffstein,Silverman_Aritimic}. Elliptic curves specified as in the equation above are said to be given in the Weierstrass normal form. When $\mathbb{K}$ coincides with the set of real numbers $\mathbb{R}$, we can graph $E(\mathbb{R})$ and view its solutions $(x,y)$ as actual points of a plane curve. An example is shown in  Fig.\ref{Fig1} (a) for a representative elliptic curve ($EC$) over the real numbers defined by the parameters $A=27$ and $B=4$. Clearly, different choices for the field $\mathbb{K}$ will result in different sets of solutions for the same cubic equation, since elliptic curves can also be regarded as particular examples of algebraic varieties. Algebraic varieties over the field of rational numbers $\mathbb{Q}$ have been investigated already by post-classical Greek mathematicians, most notably by Diophantus, who lived around 270 $CE$ in Alexandria, Egypt. In his honor, we refer to a polynomial equation in one or more variables whose solutions are sought among the integers or rational numbers as a 'Diophantine equation'. The history of Diophantine equations and elliptic curves runs central to the development of the most advanced ideas of number theory that led to the proof of the celebrated Fermat's last theorem by the British mathematician Andrew Wiles in 1995 \cite{Stewart}. 

The study of elliptic curves constitutes a major area of current research in number theory with important applications to cryptography and integer factorization.
Interestingly, when endowed with an extra point $\mathcal{O}$ at infinity, the points of elliptic curves acquire the structure of an Abelian group with the point $\mathcal{O}$ serving as the neutral group element. In particular, the group of rational points (solutions in $\mathbb{Q}$) of the elliptic curve $E(\mathbb{Q})$ is finitely generated (Mordell's theorem) and can be decomposed into the direct sum of 
$\mathbb{Z}$ with finite cyclic groups \cite{Silverman_Aritimic,Hoffstein}. More specifically, one can also show the group of rational points has the form: $E(\mathbb{Q})\cong{T}\bigoplus\mathbb{Z}^{r}$ where $T$ is a finite group consisting of torsion points (i.e., a point $P\in{E}$ satisfying $mP=\mathcal{O}$ is called a point of order $m$ in the group $E$.  All points of finite order form an Abelian subgroup called the torsion group of $E$) and $r$ is a non-negative number, called the algebraic rank of the elliptic curve $E$, which somehow characterizes its size \cite{Washington,Silverman_Aritimic}. 

An example of the composition group law for the previously introduced elliptic curve over the real numbers is illustrated in Fig.\ref{Fig1} (a) where two points with real-valued coordinates $P$ and $Q$ are summed to obtain the point $R^{\prime}$.
The simplest way to introduce the group composition law is to implement the following geometrical construction \cite{Hoffstein,Silverman_Aritimic}: we first draw the line that intersects $P$ and $Q$. This line will generally intersect the cubic at a third point, called $R$. We then define the addition $P+Q$ as the point $-R$, i.e. the point opposite $R$. It is possible to prove that this definition for addition works except in a few special cases related to the point at infinity and intersection multiplicity \cite{Hoffstein,Silverman_Aritimic}.

The type of elliptic curves that we will investigate in this paper are defined over the finite field $\mathbb{F}_p\equiv{\mathbb{Z}/p\mathbb{Z}}$
where $p$ is an odd prime number. This is the set of integers modulo $p$, which is an algebraic field when $p$ is prime.
An elliptic curve over $\mathbb{F}_p$ is still defined by equation (\ref{eq1}) where the equal sign is replaced by the congruence operation:
\begin{equation}
y^{2}\equiv{x^{3}+Ax+B}\mod{p}    
\end{equation}
where the coefficients $A,B\in\mathbb{F}_p$ and the discriminant $\Delta_{E}$ in this case must be incongruent to $0$ when reduced modulo the prime $p$.
Since $\mathbb{F}_p$ is a finite group with $p$ elements, the elliptic curve defined above has only a finite number of points that we expect to be approximately $p+1$ in number (remember the necessity to add the extra point at infinity). 
It turns out that the actual number of points $N_{p}$ of the curve $E(\mathbb{F}_p)$ fluctuates from $p+1$ within a bound $2\sqrt{p}$, which is a result proved in 1933 by Helmut Hesse. More precisely, if we define the quantity $a_{p}=p+1-N_{p}$ Hesse's theorem states that $|a_{p}|\leq{2\sqrt{p}}$ \cite{Silverman_Aritimic,Hoffstein}. 
One of the most challenging yet unsolved problems in mathematics, which is also a millennium prize problem of the Clay Mathematics Institute \cite{millenium}, is the Birch and Swinnerton-Dyer conjecture (BSD) that identifies the algebraic and the analytic rank of an elliptic curve \cite{Silverman_Aritimic,Birch}. The analytic rank of a curve $E$ is equal to the order of vanishing of the associated Dirichlet $L$-function $L(E,s)$ at $s=1$.
The $L$-function $L(E,s)$ mentioned above is a complex-valued function that is constructed based on the numbers $a_{p}$ \cite{Hasse}. This function, which is analogous to the Riemann zeta function $\zeta$ and the Dirichlet $L$-series, can be analytically continued over the whole complex plane and it encodes information on the number of solutions of $E$ modulo a prime onto the properties of the associated complex function $L(E,s)$. Moreover, $L(E,s)$ satisfies a Riemann-type functional equation connecting its values $L(E,s)$ and $L(E,2-s)$ for any $s$.
According to the Sato-Tate conjecture, the random looking fluctuations observed in the 'error term' $a_{p}$ when the prime $p$ is varied are captured by a 'sine-squared' probability distribution. This conjecture has been proved in 2008 by Richard Taylor limited to particular types of elliptic curves \cite{Taylor}.

In Fig.\ref{Fig1} (b) we show the elliptic curve over $\mathbb{F}_p$ with $p=2111$ that has the same parameters as the curve $E(\mathbb{R})$ previously shown in Fig.\ref{Fig1} (a). We note that the curve has been rescaled by a constant parameter so that the average separation between points equals 450$nm$, which enables resonant scattering responses across the visible spectrum. Apart from this irrelevant scaling, the points on this curve appear to be randomly distributed in stark contrast with its counterpart defined over the field of real numbers.  Moreover, working with  $EC$ over finite fields allows one to define the associated discrete logarithm problem that plays as essential role in elliptic curves cryptography due to its non-polynomial complexity \cite{Hoffstein,Washington}. 
Let $EC$ be an elliptic curve over $\mathbb{F}_p$ (see Fig.\ref{Fig1}(b)) and $M$ (blue circle marker) and $W$ (red circle marker) two points on the curve. The discrete logarithm problem is the problem of finding an integer $k$ such that $W=kM$. By fixing a starting point $W$ and applying this group operation repeatedly to all the points $M$ on the curve $E$ in Fig.\ref{Fig1} (b), we can obtain the point patterns shown in panels (c) and (d), which are the physical representations of the abstract discrete logarithm problem on the original curve $E$. Specifically, panels (c) and (d) display curves characterized by the coordinates $(M_x;k)$ and $(M_y;k)$ rescaled in order to have an average interparticles separtion equal to 450$nm$, respectively. These types of aperiodic deterministic structures are referred to as elliptic curve discrete logarithm ($EC$ $DL$). Clearly, the distribution of points in $EC$ $DL$ strongly depends on the choice of the initial point $W$ on the starting $EC$. In our work we have uniformly sampled $9$ starting points on $E$. We have found that the resulting $EC$ $DL$ curves can be divided into two main categories: $EC$ $DL$ point patterns that are symmetric with respect to the $x$-axis (Fig.\ref{Fig1}(c)) and others that do not show this structural symmetry and are generally less homogeneous (Fig.\ref{Fig1}(d)). Moreover, the number of elements $EC$ $DL$ cannot be controlled exactly because it depends on the value of the integer $k$. The complexity of the discrete logarithm problem for elliptic curves over finite fields is at the heart elliptic-curve cryptography (ECC), which is the most advanced approach to public-key cryptography that protects highly secure communications (up to to-secret classification) using key that are significantly smaller compared to alternative methods such as  RSA-based cryptosystems \cite{Hoffstein,Washington}. 
In what follow we will apply the statistical methods of point pattern analysis and the theory of multiple light scattering in order to investigate the structural, spectral, and scattering properties of these number-theoretic structures regarded as aperiodic photonic systems of scattering point-particles.
\begin{figure}[t!]
\centering
\includegraphics[width=12cm]{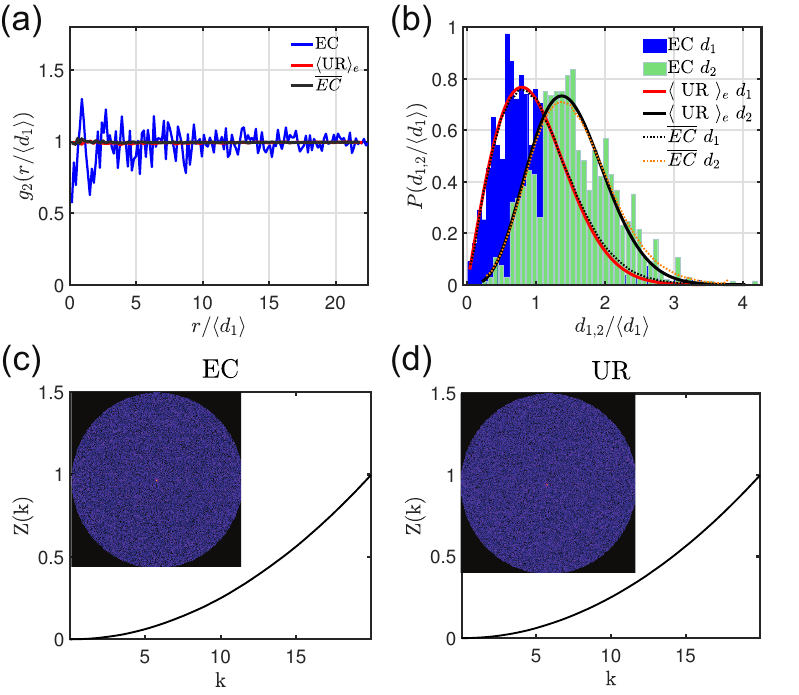}
\caption{(a) Radial distribution function $g(r)$ of the $EC$ of Fig.\ref{Fig1} (b) (blue line) as compared to the averaged two-point correlation function of 200 different disorder realizations of Poissonian point patterns (red curve). The black line identifies the averaged $g(r)$ of 900 different elliptic curves generated by all the possible combinations of the coefficients $A$ and $B$ in the range $[1,30]$. (b) First (blue bars) and second (pastel green bars) neighbor probability density function of the $EC$ point pattern generated by the equation $y^2=x^3+27x+4$ as compared to the Poissonian first and second neighbor distribution defined by Eq.(\ref{P_analitic}) \cite{Illian,SgrignuoliVogel}. The two dotted curves are the averaged $P(d_1)$ and $P(d_2)$ of the 900 different elliptic curves generated as explained above. Panels (c) and (d) display, respectively, the behavior of the integrated intensity function defined by Eq.(\ref{IIF}) of the elliptic curve of Fig.\ref{Fig1} (b) over the finite fields $\mathbb{F}_{2111}$ and of a representative Poissonian point pattern. Insets report their structure factors $S(k)$ evaluated by using Eq.(\ref{Sk}).}
\label{Fig2}
\end{figure}

\section{Structural and spectral properties of elliptic curves and discrete logarithm arrays}\label{Greensection}
In order to obtain quantitative information on the degree of local structural order of $EC$ and $EC$ $DL$ point patterns, we have computed their radial distribution functions $g(r)$, which give the probability of finding two particles separated by a distance $r$ \cite{Illian}. In Fig. \ref{Fig2} (a) we display the $g(r)$ of the representative $EC$ point pattern shown in Fig. \ref{Fig1} (b) (blue line) and we compare it with the disorder-averaged $g(r)$ of a uniform random ($UR$) structure considering 200 realizations of disorder (red line). In order to systematically analyze the correlation properties of $EC$ structures, we have considered 900 different elliptic curves generated by a uniform sample of the integer parameters $A$ and $B$ in the range [1,30]. The arithmetic average value of the $g(r)$ for all the investigated $EC$ structures is shown by the black line in Fig.(\ref{Fig2} (a)). This quantity clearly demonstrates the uncorrelated nature of $EC$ structures akin to the average behavior of Poisson random point patterns. To gain more information on the peculiar geometrical arrangements of the points compared to $UR$ systems we studied the probability density functions of the first ($P(d_1)$) and second ($P(d_2)$) neighbor distances \cite{DalNegro_Crystals, Illian}. The results of this analysis are presented in panel (b). The $P(d_1)$ and $P(d_2)$ functions can be analytically approximated using the following expression:  
\begin{equation}\label{P_analitic}
P(d_k)=\frac{2(\xi\pi r^2)^k}{r(k-1)!}~\exp(-\xi\pi r^2)
\end{equation}
where $P(d_k)$ is probability density function for the $k$-neighbor particle spacing of homogeneous Poisson point patterns with intensity $\xi$ evaluated as $N/\pi R^2$. Here $N$ is the number of points in the array, which varies between $1926$ and $2106$, and $R$ is the maximum radial coordinate of the system \cite{Illian,SgrignuoliVogel}. We notice that while the  
$P(d_1)$ and $P(d_2)$ distributions for a single $EC$ array fluctuate significantly, their average over the 900 investigated $EC$ structures with different $A$ and $B$ parameters can be precisely fitted using the analytical expression in equation (\ref{P_analitic}).

Aperiodic deterministic structures are complex structures with varying degrees of order and spatial correlations ranging from quasicrystals to more disordered amorphous structures with diffuse spectra \cite{DalNegro_Crystals,BaakeGrimm,Senechal}. In order to characterize the structural order of $EC$ point patterns,  we have analyzed their spatial Fourier spectra by evaluating the static structure factor as
\begin{equation}\label{Sk}
S(k)=\frac{1}{N}\sum_{n=1}^N\sum_{m=1}^N e^{-i\mathbf{k}\dot(\mathbf{r}_n-\mathbf{r}_m)}
\end{equation}
Interestingly, the inset of Fig.(\ref{Fig2})(c) shows that $EC$ aperiodic structures are characterized by diffuse diffraction spectra that are typically associated to homogenuous and isotropically disordered media (see the inset of Fig.\ref{Fig2} (d)). In order to get more information on the nature of these spectra, we have analyzed the behavior of the integrated intensity function defined as \cite{MaciaBook}:
\begin{equation}\label{IIF}
Z(q)=\int_{-q}^{q}\int_{-q}^{q} |S(k_x,k_y)|dk_xdk_y
\end{equation}
For two-dimensional (2D) arrays, this function characterizes the distribution of the diffracted intensity peaks contained within a square region, centered at the origin, with a maximum size of $2k\times2k$ in the reciprocal space \cite{Wang}. Interestingly, Eq.(\ref{IIF}) has also been recently used as a tool to quantitatively characterize the type of hyperuniformity of quasicrystalline point sets generated by projection method by studying its scaling behavior as $k$ tends to zero \cite{Oguz,Torquato} .
We recall that any diffraction intensity pattern can be regarded as a spectral measure $\mu_d$ that, thanks to the well-known Lebesgue's decomposition theorem \cite{Queffelec,BaakeGrimm,Senechal}, can be uniquely decomposed in term of three kinds of primitive spectral components, or a mixture of them. Specifically, any diffraction spectra measure can be expressed as $\mu_d=\mu_p\cup\mu_{sc}\cup\mu_{ac}$, where $\mu_{p}$, $\mu_{sc}$, and $\mu_{ac}$ refer to the pure-point, singular continuous, and absolutely continuous spectral components, respectively \cite{BaakeGrimm,MaciaBook,LucaBook}. In particular, in both periodic and quasiperiodic structures there are regions where $Z(q)$ is constant due to the pure point nature of the spectrum. Therefore, over spectral gap regions $Z(q)$ remains constant and presents jump discontinuities every time an isolated Bragg peak is integrated over. On the contrary, for structures with absolutely continuous Fourier spectra the integrated intensity function is continuous and differentiable. Finally, in the case of structures with singular-continuous spectra like the ones generated by the distribution of the prime number on complex quadratic fields and quaternion rings \cite{Wang}, the Bragg peaks are no longer well-separated but clustered into a hierarchy of self-similar contributions giving rise to a weak continuous component in the spectrum that smoothly increases the value of $Z(q)$ in between consecutive plateaus. In Fig.\ref{Fig2} panel (c-d) we report the calculated $Z(q)$ for $EC$ and $UR$ configurations, respectively. The results did not show any  appreciable difference compared to $UR$ systems, indicating that elliptic curves are structures characterized by an absolutely continuous diffraction spectral measure. 

We now investigate the scattering spectra and wave localization properties of the elliptic curves defined over the finite field $\mathbb{F}_{2111}$ by using the Green's matrix method. 
This formalism allows for a full description of open three-dimensional (3D) scattering resonances of large-scale structures at a relatively low computational cost if compared to traditional numerical methods such as Finite Difference Time Domain (FDTD) or Finite Element Method (FEM) techniques \cite{SgrignuoliVogel,SgrignuoliCompact,DalNegro_Crystals}. Moreover, this approach provides access to all the scattering resonances of a system composed of vector electric dipoles in vacuum and accounts for all the multiple scattering orders. The point-scatterer assumption implies that the scatterer size must be much smaller than the wavelength. Specifically, in this limit each scatterer is described by a Breit-Wigner resonance at frequency $\omega_0$ and width $\Gamma_0$ ($\Gamma_0\ll\omega_0$). The scattering resonances or quasi-modes of an array can be identified with the eigenvectors of the Green's matrix $\overleftrightarrow{G}$ which, for $N$ vector dipoles, is a $3N\times{3N}$ matrix with components \cite{SkipetrovPRL}:
\begin{equation}\label{Green}
G_{ij}=i\left(\delta_{ij}+\tilde{G}_{ij}\right)
\end{equation}
$\tilde{G}_{ij}$ has the form:
\begin{equation}\label{GreeenElectric}
\tilde{G}_{ij}=\frac{3}{2}\left(1-\delta_{ij}\right)\frac{e^{ik_0r_{ij}}}{ik_0r_{ij}}\Biggl\{\Bigl[\bm{U}-\hat{\bm{r}}_{ij}\hat{\bm{r}}_{ij}\Bigr]-\Bigl(\bm{U}-3\hat{\bm{r}}_{ij}\hat{\bm{r}}_{ij}\Bigr)\left[\frac{1}{(k_0r_{ij})^2}+\frac{1}{ik_0r_{ij}}\right]\Biggr\}
\end{equation}
 \begin{figure}[t!]
\centering
\includegraphics[width=12cm]{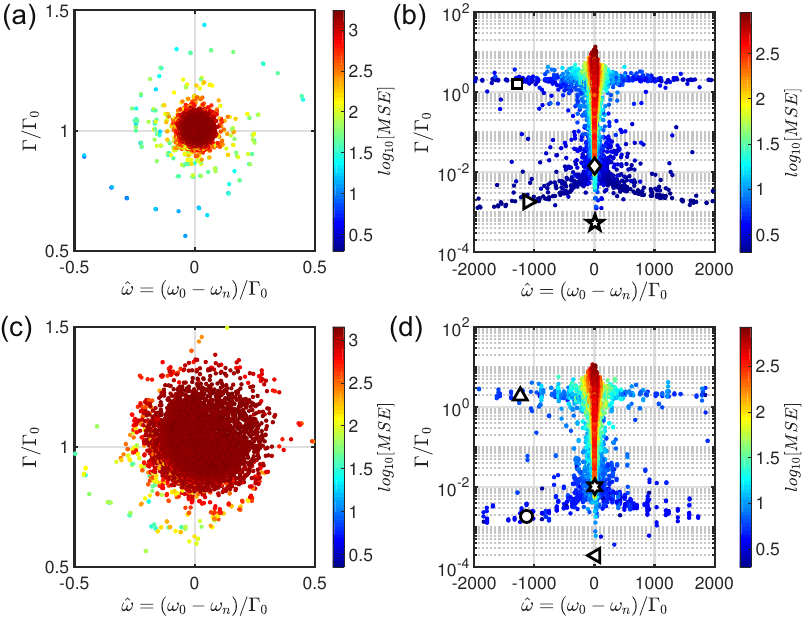}
\caption{Eigenvalues of the Green's matrix (\ref{GreeenElectric}) are shown by points on the complex plane for 1630 electric point dipoles arranged in a Poissonian configuration (panel a-b) and elliptic curve geometry (panel c-d), respectively. The $EC$ point pattern is shown in Fig.\ref{Fig1} (b). Specifically, panels (a) (c) and panels (b) (d) refer to low ($\rho\lambda^2$=0.01) and high ($\rho\lambda^2$=50) optical density, respectively. The data are colored according to the $\log_{10}$ values of the MSE. The different markers identify representative scattering resonances displayed in Fig.\ref{Fig4}. For the traditional uniform random configuration a total of at least $5\times10^4$ eigenvalues for each optical densities are considered.}
\label{Fig3}
\end{figure}
when $i\neq j$ and $0$ for $i=j$. $k_0$ is the wavevector of light, the integer indexes $i, j \in 1,\cdots,N$ refer to different particles, $\textbf{U}$ is the 3$\times$3 identity matrix, $\hat{\bm{r}}_{ij}$ is the unit vector position from the $i$-th and $j$-th scatter while $r_{ij}$ identifies its magnitude. This method is an excellent tool to study light scattered by atomic clouds but also provides fundamental insights into the physics of periodic, aperiodic, and uniform random systems of small and sufficiently well-separated scattering particles \cite{Lagendijk,Pinheiro,Pinheiro2008,Bellando,SkipetrovPRL,SkipetrovPRB,Skipetrov2015,RusekPRA,RusekPRE,Sheng,DalNegro_Crystals,Wang,SgrignuoliCompact,SgrignuoliVogel}. The Green's matrix (\ref{Green}) is a non-Hermitian matrix. As a consequence, it has complex eigenvalues $\Lambda_n$ ($n \in 1,2,\cdots, 3N$) with $\Im[\Lambda_n]=(\omega_0-\omega_n)/\Gamma_0$ and $\Re[\Lambda_n]=\Gamma_n/\Gamma_0$ \cite{RusekPRA,RusekPRE,SkipetrovPRL,SkipetrovPRB,Skipetrov2015}. 
Moreover, it is important to realize that the Green's matrix method is an eigenvalue method that captures the fundamental physics of multiple scattering of vector waves for any assembly of electric scattering point dipoles. In addition, this powerful method enables a clear separation between the geometry of the scattering arrays (the arrangement of the dipoles) and the material properties and sizes of the individual particles that are captured by a retarded polarizability or by the refractive index. As a result, the predictions of the Green's approach should be considered ``universal" in the limit of electric dipole scatterers, meaning that the size and the refractive index of the particles can be taken into account after the diagonalization of the Green's matrix by extracting the frequency $\omega_0$ and $\Gamma_0$ from the central position and the lineshape of the scattering cross section (computed using for example the Mie-Lorentz theory in the dipole limit) of a single particle. 
\begin{figure}[t!]
\centering
\includegraphics[width=\linewidth]{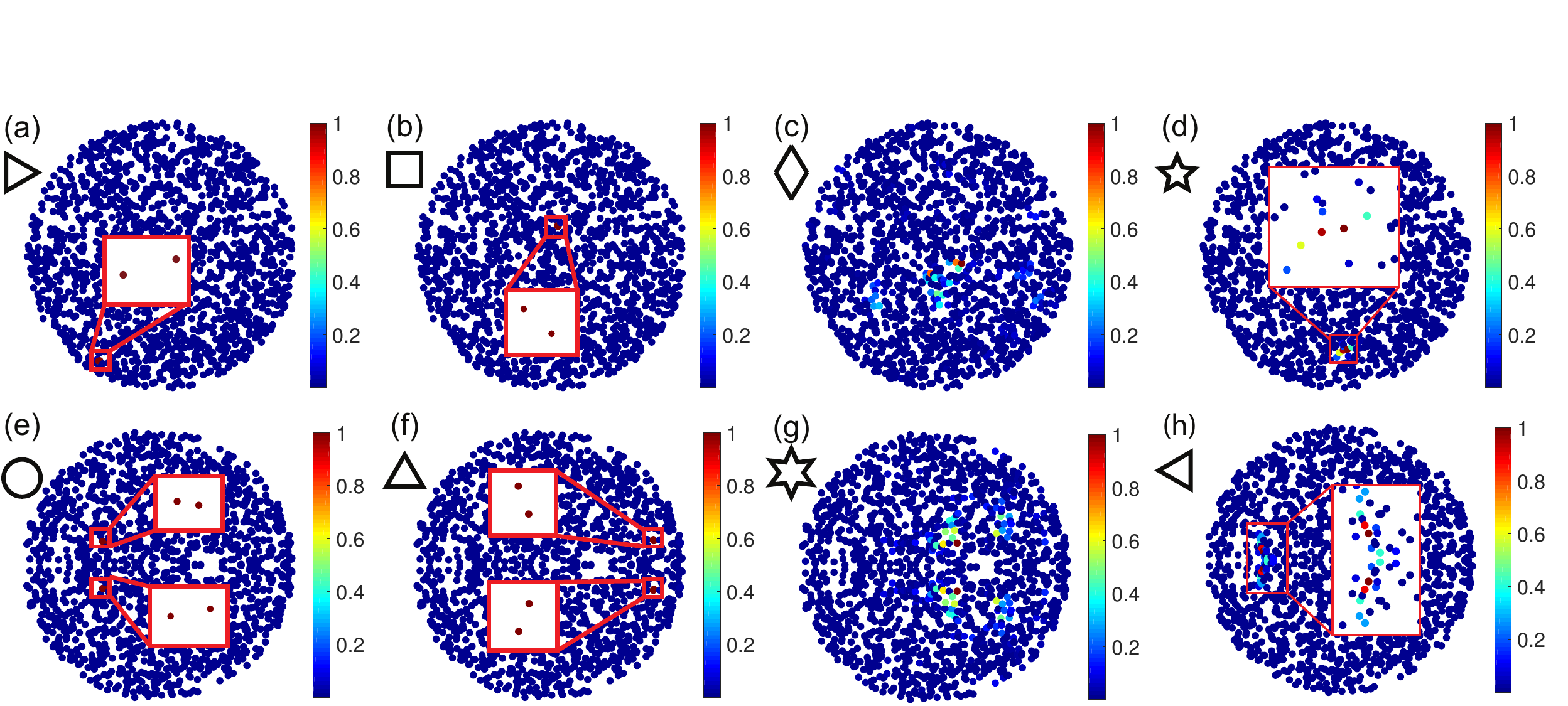}
\caption{Representative spatial distributions of the Green's matrix eigenvectors that belong to the class of scattering resonances identified in the complex plane of Fig.\ref{Fig3} (b-d). Specifically, panels (a-d) and panels (e-h) refer to the $UR$ and $EC$ configurations, respectively. }
\label{Fig4}
\end{figure}
\begin{figure}[b!]
\centering
\includegraphics[width=\linewidth]{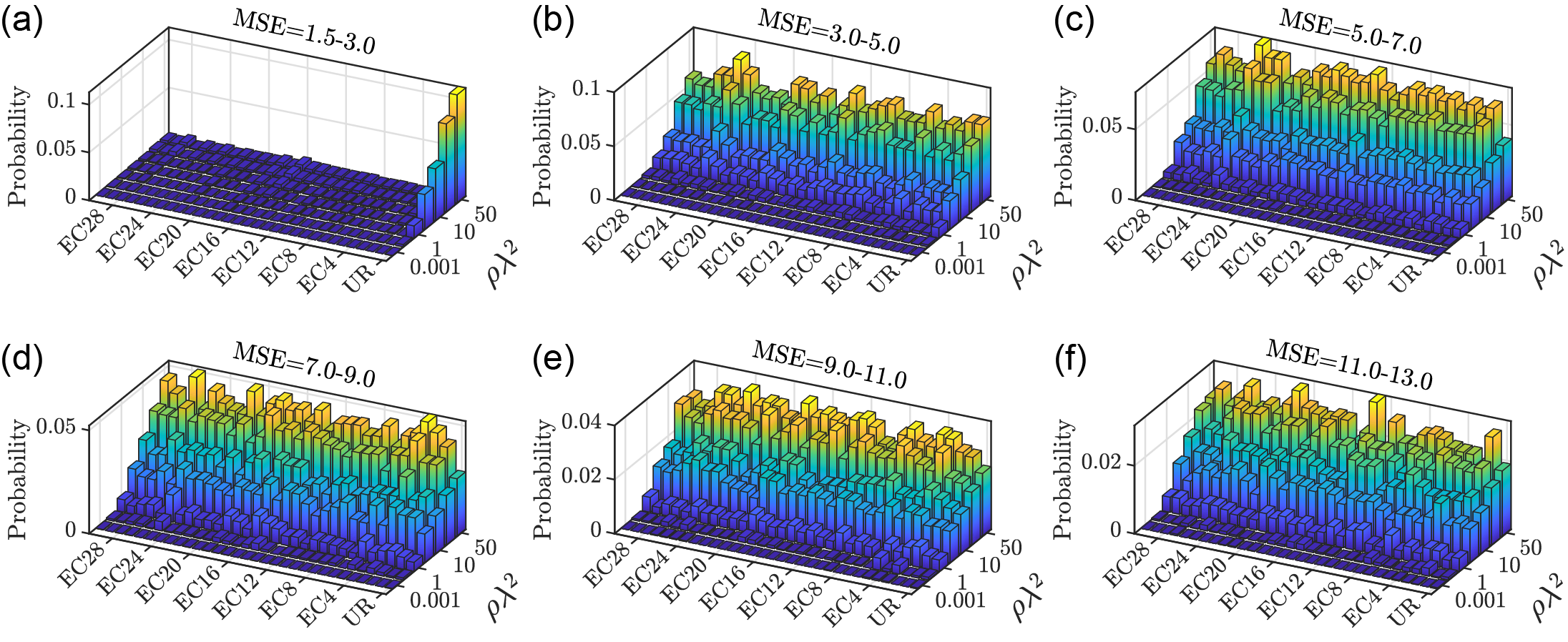}
\caption{Probability distribution of scattering resonances as a function of $\rho\lambda^2$ of representative $EC$ geometries as compared to $UR$ for various MSE intervals. Specifically, the $UR$ configuration is located in the first row of each panels, while the representative $EC$ structures, 30 planar arrays in total, are characterized by the parameters $A$ and $B$ defined by the relation $A=B=n$, where $n$ is defined in the range [1,30].}
\label{Fig5}
\end{figure}

We have applied this formalism to both $UR$ and $EC$ arrays and studied the light scattering properties in the plane of these arrays by analyzing the behavior  of their scattering resonances embedded in 3D. In order to do that, we have diagonalized numerically the $3N\times3N$ Green's matrix (\ref{GreeenElectric}). The distribution of the resonant complex poles $\Lambda_n$, color coded according to the $\log_{10}$ values of the modal spatial extent (MSE), is reported in Fig.\ref{Fig3} panels (a-b) for the $UR$ and in panels (c-d) for the representative $EC$ configuration shown in Fig.\ref{Fig1} (b), respectively. Specifically, panels (a) (c) and panels (b) (d) refer to low and high optical density $\rho\lambda^2$, respectively. Here $\rho$ is the number of particles per unit area while $\lambda$ is the optical wavelength and $\rho\lambda^2$ is a measure of the scattering strength of the systems. Instead, the MSE parameter characterizes the spatial extent of a photonic mode \cite{SgrignuoliACS}. It is important to emphasize that the dimensionality of the studied electromagnetic problem is 3D, but the electromagnetic field of a scattering resonance is not only  spatially confined in the plane of the array but it also leaks out from such a plane with a characteristic time proportional to its quality factor \cite{SgrignuoliVogel}.
\begin{figure}[b!]
\centering
\includegraphics[width=12cm]{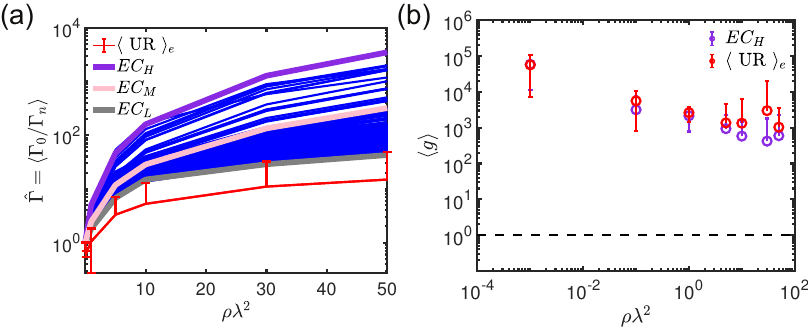}
\caption{(a) Averaged modal lifetime as a function of different optical densities of the 900 $EC$ geometries generated by all the possible combinations of the coefficients $A$ and $B$ in the range $[1,30]$ as compared to the uniform random configuration ensemble averaged over 10 different disordered realizations (red curve). The highest ($EC_H$), the middle ($EC_M$),  and the lower ($EC_L$) $\hat{\Gamma}$ trend are highlighted in violet, pink and grey color, respectively. In particular, $EC_H$, $EC_M$, and $EC_L$ are the elliptic curves over the finite field $\mathbb{F}_{2111}$ generated by the parameters combination $(a=27;b=4;)$, $(a=11;b=8;)$, and  $(a=28;b=19;)$  respectively. (b) Thouless conductance as a function of the scattering strength $\rho\lambda^2$ averaged over the frequency stripe of width 2$\Gamma_0$ centered in $\omega_0$ for the $EC_H$ (circle violet markers) and $UR$ (circle red markers) configurations, respectively. The dashed-black lines identify the threshold of the diffusion-localization transition $g=1$.}
\label{Fig6}
\end{figure}

At low optical density ($\rho\lambda^2=0.01$), the distribution of the complex poles of $N=1630$ electric point dipoles randomly located inside a circular region is highly uniform and is characterized by a circular shape with distinctive spiral arms that are weakened when the electric dipoles are arranged in $EC$ geometries (see panel (c)). These spectral regions are typically populated by scattering resonances localized over small clusters of scatterers, down to only two particles \cite{RusekPRA,Bellando}. 
The subradiant dark states, also called proximity resonances, are characterized by MSE=2 \cite{RusekPRA,Goetschy} (see also Fig.\ref{Fig4} (a-b)). Interestingly, the absence of these  scattering resonances in a class of aperiodic spirals, called Vogel spirals, was recently connected to the ability of these structures to localize vector waves thanks to their peculiar correlation properties \cite{SgrignuoliVogel}.

In order to understand the nature of the scattering resonances that characterize EC-based structures in the strong scattering regime ($\rho\lambda^2=50$), we have analyzed the spatial distributions of a few representative optical modes identified by the symbols shown in Fig.\ref{Fig3} panel (b) and (d). Fig.\ref{Fig4} shows a survey of representative spatial distribution of the Green's eigenvectors for $UR$ (panel (a-d)) and $EC$ geometries (panel (e-h)), respectively. 
From the complex eigenvalues distribution and from the selected spatial field profiles, we can clearly distinguish among three different types of scattering resonances. Let us start by analyzing the $UR$ configuration. The first type of scattering resonances correspond to short-lived quasi-modes ($\Gamma_n/\Gamma_0>1$) that are delocalized across the all arrays ($MSE\sim N$). Moreover, this spectral region is also populated by proximity resonances, as shown in panel (b). This is a clear signature of the fact that proximity resonances are not related to interference-driven light localization because they do not require multiple scattering in order to occur \cite{SgrignuoliVogel,DalNegro_Crystals,RusekPRA,Goetschy}. Finally, there are resonances that populate the dispersion branch around $\hat{\omega}=-1$ with $\Gamma/\Gamma_0<10^{-1}$. These quasi-modes are long-lived resonances with $\Gamma/\Gamma_0$ in the range $[10^{-2},10^{-1}]$ extended over almost all the particles (see Fig.\ref{Fig4} (c)) or clustered over few particles near the array boundaries (see also the discussion of Fig.\ref{Fig6} (a) for more details). However, even if the main characteristics of the complex eigenvalues distribution of $EC$ structures is very similar to the $UR$ ones, a deeper analysis unveils important differences. 
First of all, strictly speaking $EC$ structures do not show traditional proximity resonances but clustered quasi-modes ($MSE\geq4$) associated to the structural mirror symmetry along the $x$-axis, as shown in panel (e) and (f). Another important difference arises when looking at the dispersion branch around $\hat{\omega}=-1$. Indeed, the $EC$-based arrays feature longer-lived and clustered resonances ($\Gamma_n/\Gamma_0\sim10^{-4}$ with $MSE\geq$12) compared to standard $UR$ structures, see Fig.\ref{Fig4} (h). These more extended, long-lived resonances are similar to the critical scattering resonances that are typical of fractal and multifractal systems. These optical modes characterized by a power-law envelope localization and multifractal field intensity oscillations \cite{DalNegro_Crystals,Wang,Mahler,Noh,Gellermann,DalNegro_LaserPhoton}. 

To gain additional insights on the nature of the scattering resonances of $EC$-based arrays compared to $UR$ systems, we have evaluated the proportion of modes that extend over a number of particles specified by the range of the MSE values considered. In particular, we computed the  probability of the number of scattering resonances in different MSE ranges and at different optical densities. Fig.\ref{Fig5} shows the results of this study. First of all, panel (a) indicates that for $EC$ point patterns the probability of obtaining resonances localized over at most 3 scatterers is negligible compared to UR structures, shown in the first column of Fig.\ref{Fig5}. This is regardless of the value of the optical density. In contrast, proximity resonances do appear in UR systems even at low optical density ($\rho\lambda^2\geq1$). By increasing the $MSE$ threshold, the probability of finding scattering resonances localized over large clusters of particles is always larger for the analyzed $EC$ configurations compared to the UR reference structures. Therefore, our analysis provides evidence that, differently from the case of uniform random systems, the mechanism of localization in EC-based structures proceeds through wave tunneling and trapping over few-particle clusters via the formation of Efimov-like resonances \cite{Efimov}. 
\begin{figure}[b!]
\centering
\includegraphics[width=\linewidth]{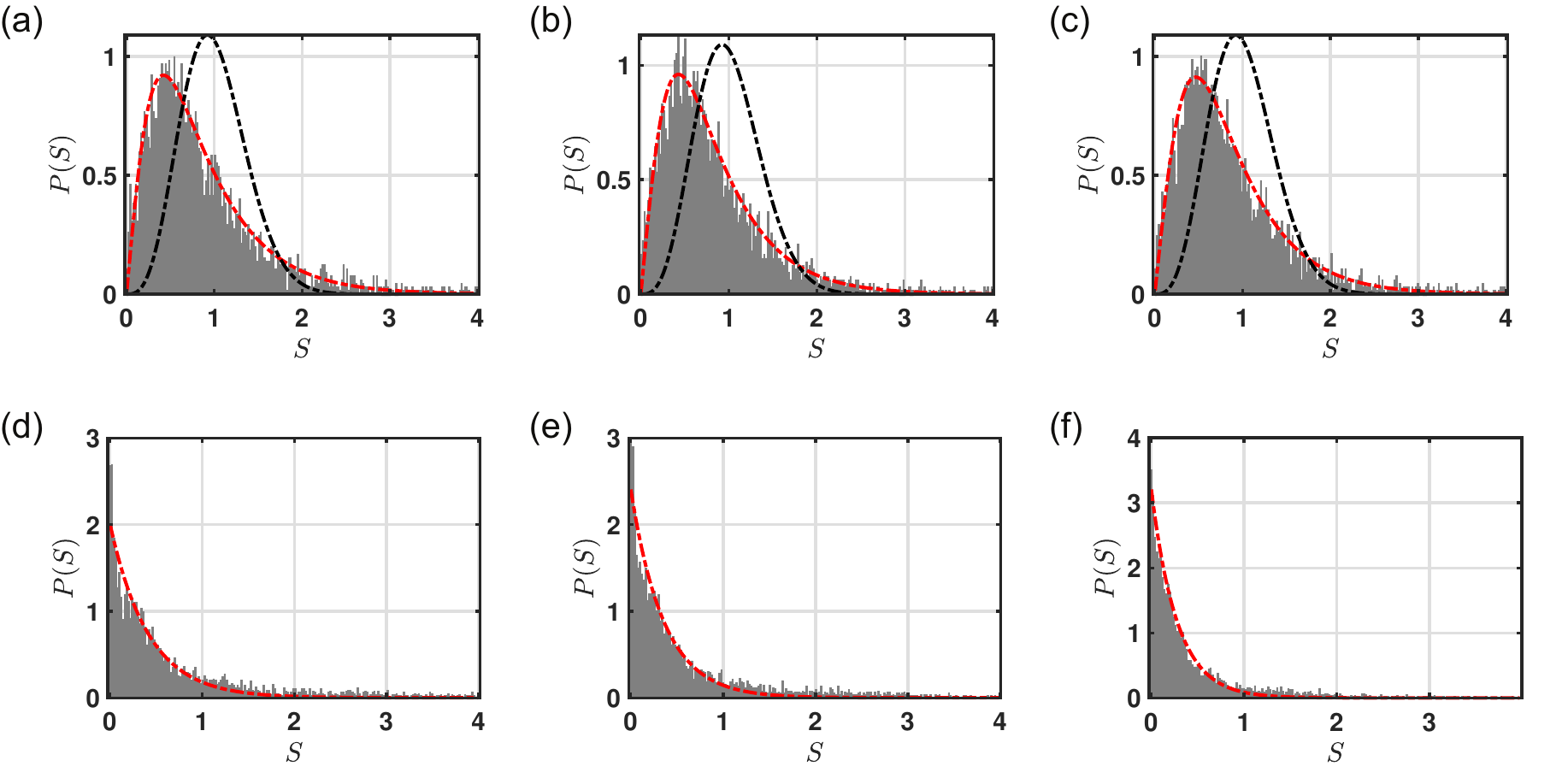}
\caption{Level spacing statistics of the Green's matrix eigenvalues for two different regimes: $\rho\lambda^2$=$0.05$ (panels (a-c)) and $\rho\lambda^2$=$50$ (panels (d-f)). Panels (a-d), (b-e), (c-f) refer to $EC_H$, $EC_M$, and $EC_L$ configurations, respectively. The fitting curves are performed by using the critical cumulative distribution \cite{DalNegro_Crystals,Wang,Zharekeshev}(dotted dashed lines in panels (a-c)) and the Poisson distribution (dotted dashed lines in panels (d-f)).The dotted dashed black lines in panels (a-c) indicates the level spacing distribution of a representative $UR$ structure defined by Eq.(\ref{Ginibre}).}
\label{Fig7}
\end{figure}

In order to analyze the light localization behavior, we have evaluated the modal average lifetime \cite{Lagendijk,Pinheiro,SgrignuoliVogel}, the Thouless conductance $g$, also called Thouless number \cite{Genack,SkipetrovIoeffe,SkipetrovSearch}, and the level spacing distribution \cite{Haake,Mehta}. 
The average modal lifetime, defined as $\hat{\Gamma}=\langle \Gamma_0/\Gamma_n\rangle$, provides the mean time that light spends inside a medium \cite{Lagendijk,Pinheiro,SgrignuoliVogel}. Fig.\ref{Fig6} (a) compares the averaged modal lifetime of the 900 different elliptic curves over $\mathbb{F}_{2111}$ with respect to the value of $\hat{\Gamma}$ produced by 10 different realization of $UR$ structures. Interestingly, all the $EC$ structures show a larger average modal lifetime for all the analyzed $\rho\lambda^2$ values demonstrating enhanced light-matter interaction compared to random uniform random systems.
The ability to confine and eventually localized light is also described by the Thouless conductance, which is a parameter that characterizes the degree of spectral overlap between different optical scattering resonances. In order to demonstrate light localization, the Thouless conductance, which is proportional to the scattering mean free path, must decrease below the value $1$ when increasing the scattering strength, $i.e.$ increasing the optical density $\rho\lambda^2$. Within the Green's matrix formalism, it is defined as the ratio of the dimensionless lifetime $(\delta\omega)^{-1}=1/\Im[\Lambda_{n}]$ to the spacing of nearest dimensionless resonance frequencies $\delta\omega=\Re[\Lambda_{n}]-\Re[\Lambda_{n-1}]$\cite{SkipetrovPRL}:
\begin{equation}\label{Conductance_g}
\langle g\rangle=\frac{\overline{\delta\omega}}{\overline{\Delta\omega}}=\frac{(\overline{1/\Im[\Lambda_{n}]})^{-1}}{\overline{\Re[\Lambda_{n}]-\Re[\Lambda_{n-1}]}}
\end{equation}
where $\overline{\{\cdots\}}$ indicates the average of $g$ over a frequency interval of width 2$\Gamma_0$ centered in $\omega_0$. This frequency stripe selection is necessary due to the strong frequency dependence of the light localization behavior \cite{SgrignuoliVogel}. On the other hand, averaging over all scattering frequencies will produce biased results due to mixing of different types of light regimes \cite{SkipetrovSearch}. Differently from the uniform random media, we do not need to consider any ensemble averages because the $EC$ structures are deterministic. Fig.\ref{Fig6} (b) compares the semilog plot of the Thouless conductance, as a function of $\rho\lambda^2$, obtained by using Eq.(\ref{Conductance_g}) after diagonalizing the $3N\times3N$ Green's matrix of the $EC$ structure with the highest $\hat{\Gamma}$ (violet line in Fig.\ref{Fig6} (a)) with respect to the uniform random scenario. Even though $EC$ structures are characterized by longer-lived critical resonances than $UR$ systems, the $\langle g\rangle$ parameter clearly indicates that the light localization transition is never achieved, since $\langle g\rangle$ is always larger than $1$. This is due to two factors: the presence of degenerate proximity resonances (like the ones shown in Fig.\ref{Fig4} (e-f)) and the absence of any structural correlations. The absence of structural correlations was recently identified as the factor preventing light localization to occur in uniform random arrays when the vector nature of light is taken into account \cite{SgrignuoliVogel,Skipetrov2015,SkipetrovPRL}. 
\begin{figure}[t!]
\centering
\includegraphics[width=12cm]{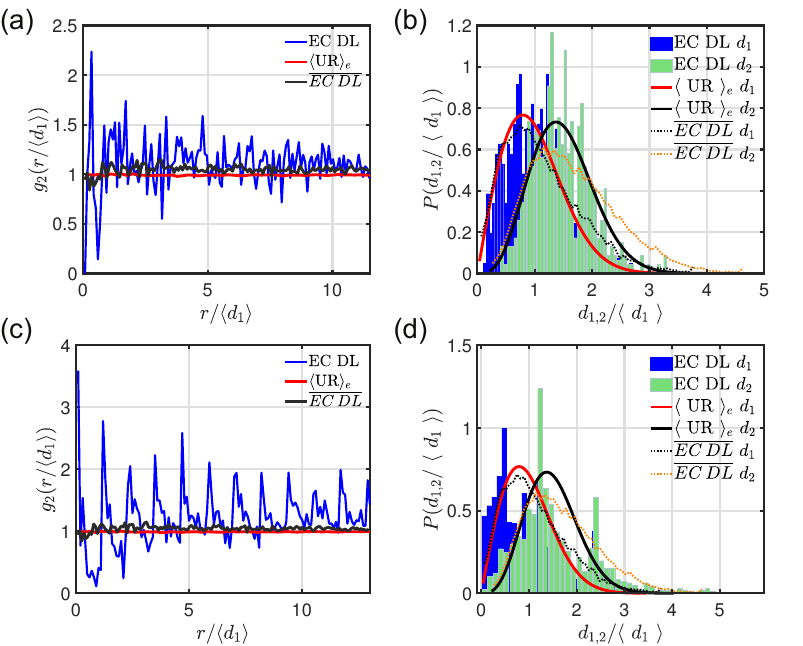}
\caption{Panels (a-b) and panels (c-d) report the radial distribution function $g(r)$ and the first and second neighbor probability density function of the $EC$ $DL$ point patterns reported in Fig.\ref{Fig1} panels (c) and (d), respectively. Moreover, their averaged values, with respect to 72 different $EC$ $DL$ geometries (generated by randomly selecting the starting point $W$ from the elliptic curve point patterns generated by the coefficients combination $(a=27;b=4)$ and $(a=28;b=19)$, named $EC_H$ and $EC_L$ respectively), are compared with respect the $UR$ scenario.}
\label{Fig8}
\end{figure}

In order to further investigate the spectral properties of $EC$ arrays we considered the distribution of level spacing $P(s)$ that provides important information about the electromagnetic propagation for both closed- and open-scattering systems \cite{DalNegro_Crystals}. Indeed, the shape of $P(s)$ depends on the spatial extent of the system eigenmodes. In particular, for open weakly disordered random media the probability density function of spacings between nearest eigenvalues $\Lambda_i$ and $\Lambda_{i+1}$ is described by:
\begin{equation}\label{Ginibre}
P(s)=\frac{3^4\pi^2}{2^7}s^3\exp\left(-\frac{3^2\pi}{2^4}s^2\right)
\end{equation}
where $s=|\Delta\Lambda/\langle|\Delta\Lambda|\rangle|$ is the normalized eigenvalue spacing \cite{Haake,Mehta,Skipetrov2015}. The important feature of this equation is the so-called level-repulsion phenomenon: $P(s)\rightarrow0$ when $s\rightarrow0$. The level repulsion is a characteristic of extended/delocalized scattering resonances that repel each other in the complex plane \cite{DalNegro_Crystals,Skipetrov2015}. On the contrary, the appearance of localized states leads to a suppression of the eigenvalue repulsion because two spatially localized states hardly influence each other when strongly localized in different parts of the medium. Consequently, distinct modes with infinitely close energies are allowed and the distribution of level spacings is described in this more localized regime by the Poisson distribution:
\begin{equation}\label{Poisson}
P(s)\approx\exp{(-s)}
\end{equation}
Notably, the level spacing statistics is very well described by Eq.(\ref{Poisson}) in the strong scattering regime for closed as well as for open (dissipative) systems \cite{DalNegro_Crystals,Haake}. 
\begin{figure}[t!]
\centering
\includegraphics[width=\linewidth]{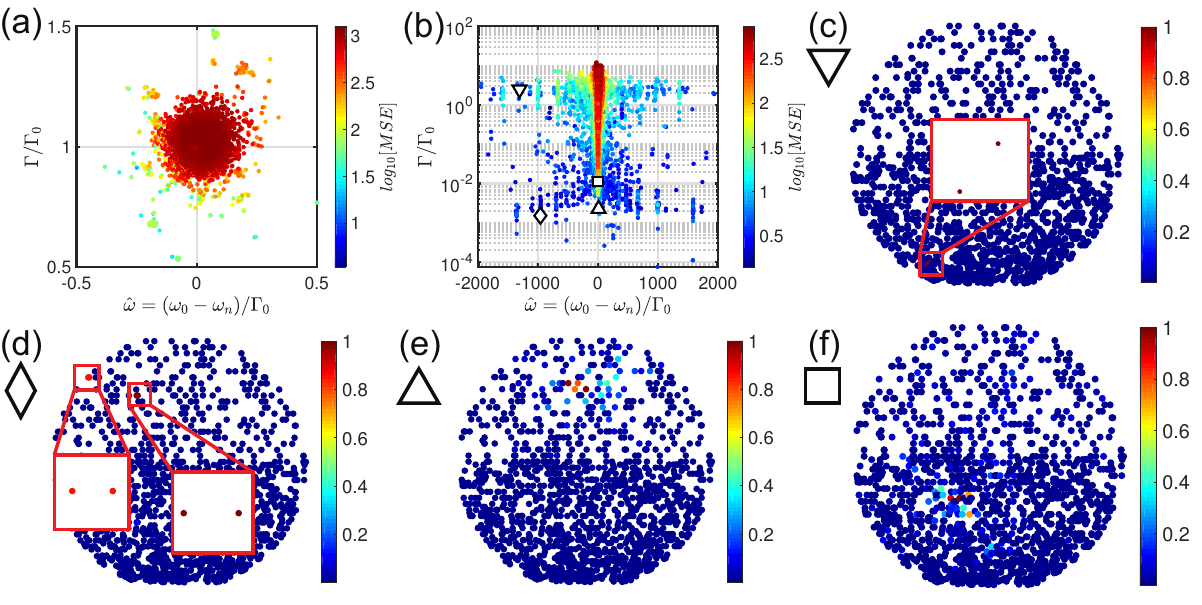}
\caption{Panels (a) and (b) display the complex eigenvalues distributions of a representative $EC$ $DL$ configuration (integer starting point equal to $W=(379;1735)$ on the elliptic curve defined by the equation $y^2=x^3+27x+4$ defined over the finite field $\mathbb{F}_{2111}$) of the Green's matrix (\ref{GreeenElectric}) for two different optical densities, respectively. The data are colored according to the $\log_{10}$ values of the MSE. The different markers in panel (b) identify representative scattering resonances displayed in panels (c-f). In particular, panels (c-d) display , respectively, a proximity and a clustered scattering resonance on 4 particles, while panels (e-f) show two representative modes with the lowest decay rates.}
\label{Fig9}
\end{figure}

However, the level statistics of deterministic aperiodic systems displays different features with respect to the uniform random scenario. In our previous works, we have investigated the transition from the presence to the absence of level repulsion by increasing $\rho\lambda^2$ in different open, deterministic, and aperiodic planar systems \cite{DalNegro_Crystals, Wang,SgrignuoliVogel}. We have found that the distribution obeys at $\rho\lambda^2$<1 the critical cumulative probability density:
\begin{equation}\label{CriticalDistribution}
I(s)=\exp\Biggl[\mu-\sqrt{\mu^2+(A_cs)^2}\Biggl]
\end{equation} 
where $\mu$ and $A_c$ are fitting parameters. This is attributed to the formation of a large number of critical scattering resonances. Indeed, Eq.(\ref{CriticalDistribution}) was successfully applied to describe the energy level spacing distribution of an Anderson Hamiltonian containing $10^6$ lattice sites at the critical disorder value, $i.e.$ at the metal-insulator threshold where it is known that all the wave functions exhibit multifractal scaling properties \cite{Zharekeshev}. We remark that the presence of a critical statistics in the spectral behavior of $EC$ structures occurs over a broad range of optical densities compared to the case of random media in which criticality is achieved only at the threshold density $\rho_c$ \cite{DalNegro_Crystals, Wang,SgrignuoliVogel}. 

The results shown in Fig.\ref{Fig7} demonstrate that the critical behavior is a generic attribute of all the investigated $EC$ structures. Indeed, panels (a-d), (b-e), (c-f) show a transition from the presence to the absence of level repulsion by increasing $\rho\lambda^2$ for the $EC_H$, $EC_M$, and $EC_L$ point patterns, defined in Fig.\ref{Fig6} (a), respectively. At low optical density (panels (a-c)), the $P(s)$ of the $EC$ point patterns and $UR$ configurations are well described by, respectively, Eq.(\ref{CriticalDistribution}) (dotted dashed red lines) and Eq.(\ref{Ginibre}) (dotted dashed black lines). On the other hand, $P(s)$ follows the Poisson distribution (\ref{Poisson}) for both $UR$ and $EC$ configurations for high optical density ($\rho\lambda^2=50$). Since the presence of a critical statistics is associated to the multifractal nature of the spectrum, the criticality discovered in $EC$ structures opens intriguing opportunities to engineer wave transport in these novel aperiodic systems \cite{DalNegroScirep,ChenScirep}.
\begin{figure}[t!]
\centering
\includegraphics[width=\linewidth]{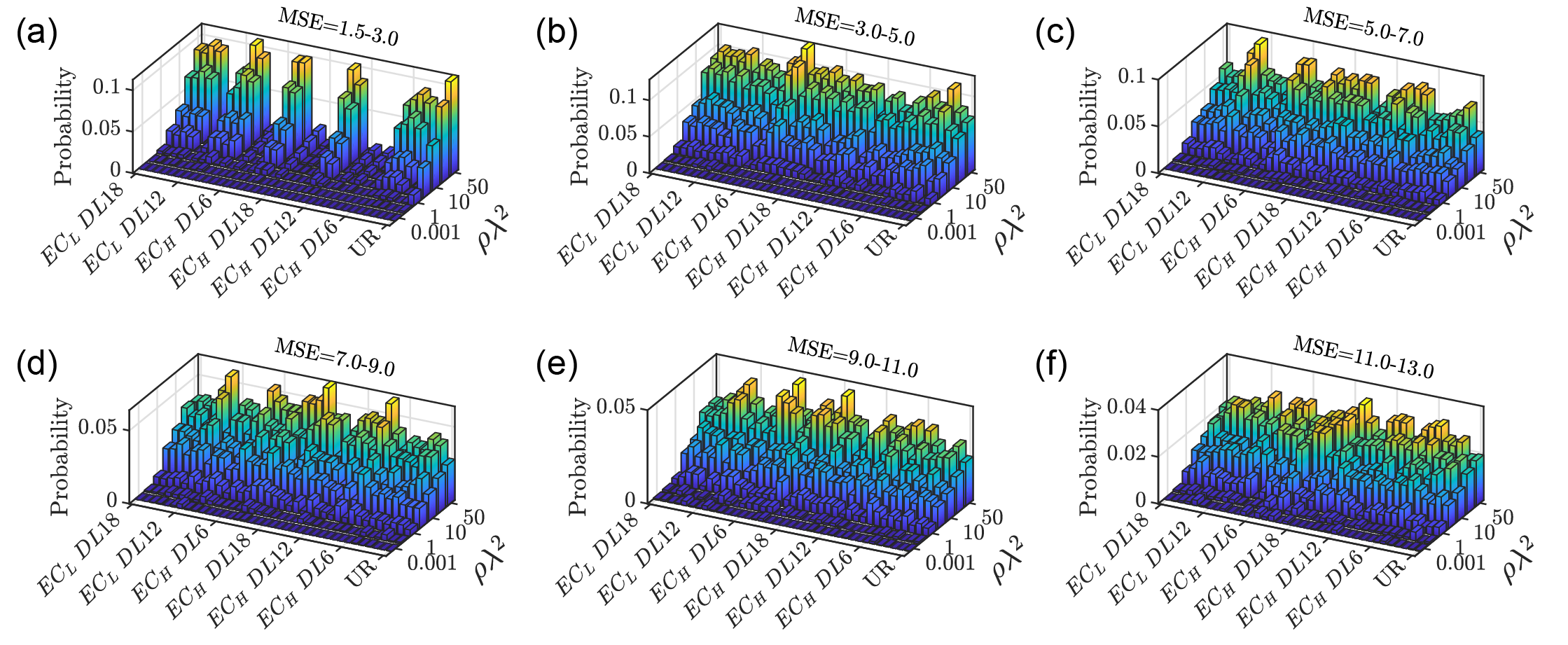}
\caption{Probability distribution of scattering resonances as a function of $\rho\lambda^2$ of 36 representative $EC$ $DL$ geometries as compared to $UR$ structures for various MSE intervals. Specifically, the $UR$ configuration is reported in the first row of each panels. The reported $EC$ $DL$ structures are generated by randomly selecting 9 integers starting points $W$ on the two elliptic curves defined by the parameter combination $(a=27;b=4;)$, and $(a=28;b=19;)$ over the finite field $\mathbb{F}_{2111}$, respectively. These two $EC$ point patterns are the $EC_H$ and $EC_L$ structures introduced in Fig.\ref{Fig6}. Each selected integer $W$ generates two different $EC$ $DL$ aperiodic point patterns by solving the two discrete logarithmic problems $W=kM_x$ and $W=kM_y$, where $M_x$ and $M_y$ are the components of a each point $M$ of $EC$ that satisfies these two relations (see  \ref{ECDL_implementation} and in the caption of Fig.\ref{Fig1} for more details). In particular, the 9 integer points selected on the $EC_H$ are: (1893;1826), (114;1753), (375;1739), (340;936), (1124;999), (1881;1246), (1902;389), (1129;395), and (305;329). On the other hand, the 9 integer points selected on the $EC_L$ are: (1719;1909), (1122;1836), (382;1761), (212;889), (1021;1138), (1841;1105), (1768;330), (1066;243), and (295;235).}
\label{Fig10}
\end{figure}

We now address the structural and spectral properties of aperiodic point patterns obtained by the solution of the discrete logarithm problem, as discussed in Section \ref{ECDL_implementation}. Fig.\ref{Fig8} displays the main results of the structural analysis, based on the radial distribution function and on the first and second neighbor distributions. $EC$ $DL$ structures, generated by the coordinate $(M_y;k)$, show higher degree of structural correlations than the $EC$ $DL$ that are symmetric with respect to the $x$-axis, $i.e$ produced by the pairs $(M_x;k)$. However, after averaging over 72 different $EC$ $DL$ arrays generated by randomly selecting the starting point $W$ from the $EC_H$ and $EC_L$ point patterns (the elliptic curve configurations that show the highest and lowest modal lifetime, as reported in Fig.\ref{Fig6} (a)), the $g(r)$ becomes constant and very close to $1$ in value, indicating absence of any structural correlations (see the black lines in Fig.\ref{Fig8} panel (a) and (c)). Moreover, also the averaged first (black dash dotted lines in panels (b) and (d)) and second (orange dash dotted lines in panels (b) and (d)) neighbor distributions are very similar to the analytical expression of  Eq.(\ref{P_analitic}) valid for homogeneous Poisson point processes \cite{Illian}. Therefore, we have found that on average also the $EC$ $DL$ structures are spatially uncorrelated point patterns (incidentally, this property explains why the discrete logarithm problem on elliptic curve is a very hard problem). This behavior is also confirmed by analyzing their spectral properties via the diagonalization of the matrix (\ref{GreeenElectric}). Indeed, the results of Fig.\ref{Fig9} are very similar to the ones reported in Fig.\ref{Fig3} for the EC structures. Specifically, the complex eigenvalues distribution of $EC$ $DL$ point patterns at low optical density ($\rho\lambda^2=0.01$) shows the same characteristics of elliptic curves: a circular disk region as for the $UR$ structures but without the distinctive spiral arms populated by the proximity resonances (see Fig.\ref{Fig9} (a)). Instead, the distribution of the complex scattering poles at large optical density shows similar features in both $EC$ $DL$ and $UR$ arrays. In particular, both proximity and clustered quasi-modes, with $MSE\geq4$, are present. Proximity resonances populate mostly the spectral region with $\Gamma_n>\Gamma_0$, while these clustered optical modes characterize the sub-radiant spiral arms (see Fig.\ref{Fig9} (c-d)). Moreover, the dispersion branch around $\hat{\omega}=-1$ is characterized by both scattering resonances clustered on few particles near the array boundaries, similar to the $UR$ scenario, and by critical quasi-modes, as shown in Fig.\ref{Fig9} panels (e) and (f), respectively. Indeed, a clear spectral region characterized by longer-lived modes with large value of $MSE$ is clearly visible is Fig.\ref{Fig9} (b) when $\hat{\omega}=-1$ and $\Gamma/\Gamma_0<10^{-1}$. 

\begin{figure}[t!]
\centering
\includegraphics[width=12cm]{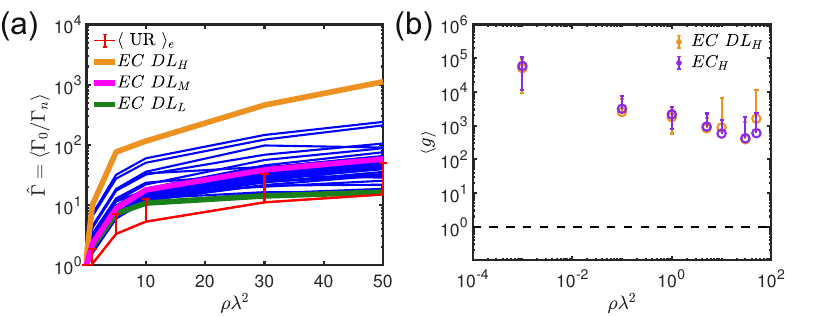}
\caption{(a) Averaged modal lifetime as a function of different optical densities of 36 $EC$ $DL$ geometries generated by randomly selecting the point $W$ from the $EC_H$ and $EC_L$ point patterns as compared to the uniform random configuration ensemble averaged over 10 different disordered realizations (red line). The highest ($EC$ $DL_H$), the middle ($EC$ $DL_M$),  and the lower ($EC$  $DL_L$) $\hat{\Gamma}$ trends are highlighted in orange, cyan, and green colors, respectively. Specifically, $EC$ $DL_H$ and $EC$ $DL_M$ are the point patterns characterized by the coordinates $(M_x;k)$ and $(M_y;k)$ generated, respectively, by solving the discrete logarithmic problem $W=kM$ associated to the $EC_H$ when $W$ is equal to (375;1739) and (1902;389). $EC$ $DL_L$ is, instead, generated by solving the discrete logarithmic problem $W=kM_x$ associated to the elliptic curve $y^2=x^3+28x+19$ defined over the finite field $\mathbb{F}_{2111}$ when the integer $W$ is equal to (295;235). (b) Thouless conductance as a function of the scattering strength $\rho\lambda^2$ averaged over the frequency stripe of width of 2$\Gamma_0$ centered in $\omega_0$ for the $EC$ $DL_H$ (circle carrot orange markers) and $EC_H$ (circle violet markers) configurations, respectively. The dashed-black lines identify the threshold of the diffusion-localization transition $g=1$.}
\label{Fig11}
\end{figure}

\begin{figure}[t!]
\centering
\includegraphics[width=\linewidth]{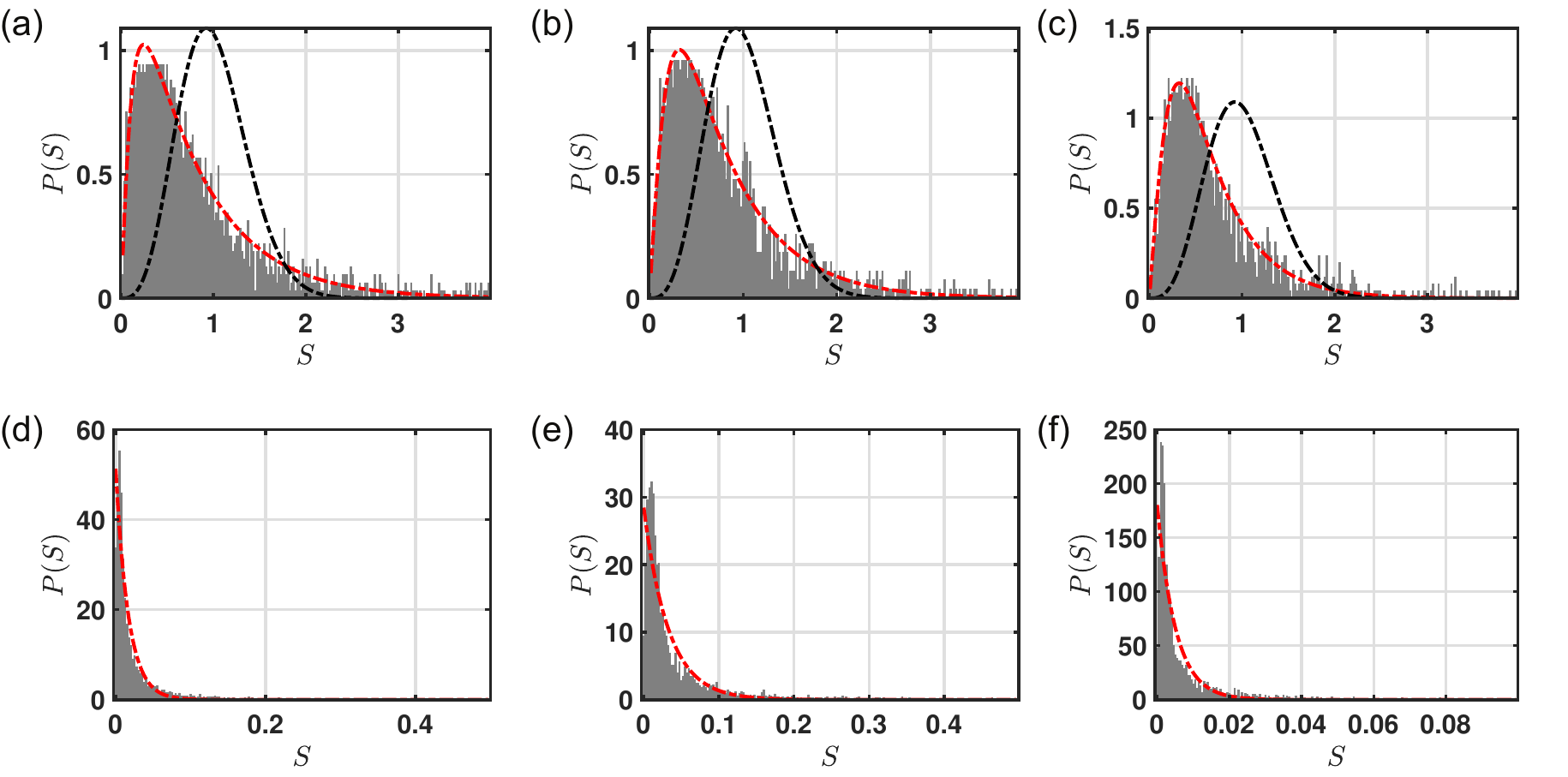}
\caption{Level spacing statistics of the Green's matrix eigenvalues for two different regimes: $\rho\lambda^2$=$0.05$ (panels (a-c)) and $\rho\lambda^2$=$50$ (panels (d-f)). Panels (a-d), (b-e), (c-f) refer to $EC$ $DL_H$, $EC$ $DL_M$, and $EC$ $DL_L$ configurations, respectively. The fitting curves are performed by using the critical cumulative distribution \cite{DalNegro_Crystals,Wang,Zharekeshev}(dotted dashed lines in panels (a-c)) and the Poisson distribution (dotted dashed lines in panels (d-f)). The dotted dashed black lines in panels (a-c) indicates the level spacing distribution of a representative $UR$ structure defined by Eq.(\ref{Ginibre})}
\label{Fig12}
\end{figure}

In order to understand how critical quasi-modes influence the light-matter interaction properties of $EC$ $DL$ geometry, we have also evaluated their probability density function by selecting different $MSE$ ranges. We discovered that the probability of finding scattering resonances localized over a clusters of scatterers is always larger than in the $UR$ scenario (see Fig.\ref{Fig10}). Of particular interest is the situation depicted in Fig.\ref{Fig10} (a) where the $MSE$ range is fixed between 1.5 and 3, $i.e$ the resonances are localized over at most 3 particles. Whereas $EC$ point patterns are always characterized by the absence of proximity resonances, $EC$ $DL$ structures instead show the presence of long-lived, strongly localized sub-radiant states even for the lowest $MSE$ range considered. Specifically, we observed that proximity resonances are always present in the type of $EC$ $DL$ point patterns that lack reflection symmetry, discussed in section \ref{ECDL_implementation}. Moreover, the modal average lifetime (Fig.\ref{Fig11} (a)), the Thouless conductance (Fig.\ref{Fig11} (b)), and the level spacing statistics (Fig.\ref{Fig12}) clearly demonstrate the role played by the critical scattering resonances also for the case of $EC$ $DL$ structures. However, we found that the average modal lifetime of 36 $EC$ $DL$ aperiodic structures is always larger than the $UR$ scenario. Moreover, the Thouless conductance of the $EC$ $DL$ and $EC$ structures with the largest $\hat{\Gamma}$ (orange and violet line in Fig.\ref{Fig6} (a) and Fig.\ref{Fig11} (a), respectively) are comparable and both larger than what can be achieved in $UR$ systems, demonstrating the potential to obtain stronger light-matter interaction in these novel aperiodic arrays. 

\section{Light scattering properties and the extended Green's matrix method}
As discussed in the previous section, the Green's matrix spectral method is an excellent approximation to study light scattering by atomic clouds and can give fundamental insights into the physics of multiple scattered light by small particles within the dipole approximation. However, this method is an oversimplification in the case of realistic scatterers that are, instead, characterized by higher-order multipolar resonances. The number of these peaks depends by the scatterer material and by the size parameter $x$ defined as $kR$, where $k$ is the wavelength number, while $R$ is the scatterer radius. Specifically, light scattering by a homogeneous, isotropic and spherical particle with radius $R$ illuminated by a linearly polarized plane wave traveling in the z-direction $\textbf{k}=[0;0;k]=[0;0;\omega/c]$ can be calculated by using the Mie-Lorentz theory \cite{Bohren}. If the size of scatterers in an array is smaller than the incident wavelength and they are far enough from each others, the light scattering problem can be described by using only the dipolar term in the general multipolar expansion \cite{Nieto}. However, the interplay between the electric and magnetic dipolar responses of small particles is a key ingredient in determining their directional scattering features. Therefore, in order to obtain a more realistic description of light scattering from these complex arrays we must go beyond the simple electric dipole framework. For this reason, we provide an extension of the Green's matrix method that takes into account both the first-order Mie-Lorentz coefficients, referred to as the electric and magnetic coupled dipole approximation (EMCDA) \cite{Bohren,Camara}. In this approximation, each particle is characterized by two dipoles (electric dipole (ED) and magnetic dipole (MD)) corresponding to the induced electric and a magnetic polarizability \cite{Yurkin}. 

In order to derive rigorously the EMCDA approximation, we must start from the electric $\bm{E}_p$ and magnetic fields $\bm{H}_p$ at a distance $r$ and direction $\bm{n}$ produced by an electric dipole $\textbf{p}$. In the \emph{cgs} unit system, we have \cite{Eyges}:
\begin{eqnarray}\label{Ep}
\begin{aligned}
&\bm{E}_p=\left[\bm{p}\frac{e^{ikr}}{r}\left(k^2-\frac{1}{r^2}+\frac{ik}{r}\right)+\bm{n}(\bm{n}\cdot\bm{p})\frac{e^{ikr}}{r}\left(-k^2+\frac{3}{r^2}-\frac{3ik}{r}\right)\right]\\
&\bm{H}_p=\left[(\bm{n}\times\bm{p})\frac{e^{ikr}}{r}\left(k^2+\frac{ik}{r}\right)\right]\label{Hp}
\end{aligned}
\end{eqnarray}
Equivalently, the field $\bm{E}_m$ and $\bm{H}_m$ produced by a magnetic dipole $\bm{m}$ are give by \cite{Eyges}
\begin{eqnarray}\label{Em}
\begin{aligned}
&\bm{E}_m=-(\bm{n}\times\bm{m})\frac{e^{ikr}}{r}\left(k^2+\frac{ik}{r}\right)\\
&\bm{H}_m=\left[\bm{m}\frac{e^{ikr}}{r}\left(k^2-\frac{1}{r^2}+\frac{ik}{r}\right)+\bm{n}(\bm{n}\cdot\bm{m})\frac{e^{ikr}}{r}\left(-k^2+\frac{3}{r^2}-\frac{3ik}{r}\right)\right]\label{Hm}
\end{aligned}
\end{eqnarray}
By introducing the coefficients $a$, $b$, and $d$ defined as \cite{Bohren,Camara}
\begin{equation}\label{aij}
\begin{aligned}
 a=\frac{e^{ik_0r}}{r}k_0^2 \left(1-\frac{1}{k_0^2r^2}-\frac{1}{ik_0r}\right)
\end{aligned}
\end{equation}
\begin{equation}\label{bij}
\begin{aligned}
 b=\frac{e^{ik_0r}}{r}k_0^2 \left(-1+\frac{3}{k_0^2r^2}+\frac{3}{ik_0r}\right)
\end{aligned}
\end{equation}
\begin{equation}\label{dij}
\begin{aligned}
  d=\frac{e^{ik_0r}}{r}k_0^2\left(1 - \frac{1}{i k_0r}\right)
\end{aligned}
\end{equation}
we can rewrite Eq.(\ref{Ep}-\ref{Hm}) in a shorter notation:
\begin{eqnarray}\label{ShortNot}
\begin{aligned}
&\bm{E}_p=\bm{p}~a+\bm{n}(\bm{n}\cdot\bm{p})~b\\
&\bm{H}_p=(\bm{n}\times\bm{p})~d\\
&\bm{E}_m=-(\bm{n}\times\bm{m})~d\\
&\bm{H}_m=\left[\bm{m}~a+\bm{n}(\bm{n}\cdot\bm{m})~b\right]
\end{aligned}
\end{eqnarray}

As a first step, our goal is to evaluate the total electric and magnetic fields at the $ith$ particle ($\bm{E}_{i}$ and $\bm{H}_{i}$) resulting from the electric and magnetic dipole moments of the $jth$ particle. Explicitly, we can write, by using Eq.(\ref{ShortNot}), $\bm{E}_{i}$ and $\bm{H}_{i}$ as:
\begin{equation}\label{EM_i}
\begin{aligned}
\bm{E}_i=\bm{E}^{j}_p+\bm{E}^{j}_m=\left[a_{ij}~\bm{p}_j+b_{ij}~(\bm{n}_{ij}\cdot\bm{p}_j)\bm{n}_{ij}\right]-d_{ij} ~(\bm{n}_{ij}\times\bm{m}_{j})\\
\bm{H}_i=\bm{H}^{j}_p+\bm{H}^{j}_m=d_{ij}~(\bm{n}_{ij}\times\bm{p}_{j})+[a_{ij}~\bm{m}_j+b_{ij}~(\bm{n}_{ij}\cdot\bm{m}_j)\bm{n}_{ij}]
\end{aligned}
\end{equation}
where the electric and magnetic dipole moments at the $jth$ particle position are defined as $\bm{p}_j=\alpha_E\bm{E}_j$ and $\bm{m}_j=\alpha_H\bm{H}_j$, respectively \cite{Chaumet}. The electric and magnetic polarizabilities $\alpha_E$ and $\alpha_M$ (that have units of a volume) are related to the first order Mie-Lorentz coefficients $a_1$ and $b_1$ as \cite{Yurkin,Doyle}:
\begin{eqnarray}\label{polarizability}
\begin{aligned}
  &\alpha_E=\frac{3\pi i}{2k_0^3} a_1\\
  &\alpha_H=\frac{3\pi i}{2k_0^3} b_1
\end{aligned}
\end{eqnarray}
Here, $k_0$ is the wavenumber of the background medium, while $a_1$ and $b_1$ are derived from the equation
\begin{eqnarray}\label{a}
\begin{aligned}
  &a_{\nu}=\frac{n\psi_{\nu}(nkR)\psi_{\nu}'(kR)-\psi_{\nu}(kR)\psi_{\nu}'(nkR)}{n\psi_{\nu}(nkR)\xi_{\nu}'(kR)-\xi_{\nu}(kR)\psi_{\nu}'(nkR)}\\
  &b_{\nu}=\frac{\psi_{\nu}(nkR)\psi_{\nu}'(kR)-n\psi_{\nu}(kR)\psi_{\nu}'(nkR)}{\psi_{\nu}(nkR)\xi_{\nu}'(kR)-n\xi_{\nu}(kR)\psi_{\nu}'(nkR)} \label{b}
\end{aligned}
\end{eqnarray}
where $R$ is the radius of the spherical scatterer, and $n$ is the relative refractive index of the nanosphere with respect to the background medium. $\psi_{\nu}(x)$ and $\xi_{\nu}(x)$ are the Riccati-Bessel functions constructed from spherical Bessel functions via $\psi_{\nu}(x)=xj_{\nu}(x)$ and $\xi_{\nu}(x)=xh_{\nu}^{(1)}(x)$. In addition, $j_{\nu}(x)$ is the spherical Bessel function of the first type, and $h_{\nu}^{(1)}(x)$ is the spherical Hankel function of the first type. By substituting the dipole moments expressions into Eq.(\ref{EM_i}), we finally obtain the total electric and magnetic fields at the $ith$ particle in the form: 
\begin{equation}\label{Ei}
  \textbf{E}_{i} = a_{ij}\alpha_{E}\textbf{E}_{j} + b_{ij}\alpha_{E}(\textbf{E}_{j}\cdot\textbf{n}_{ji})\textbf{n}_{ji} -d_{ij}\alpha_{H}(\textbf{n}_{ji}\times\textbf{H}_{j})
  \end{equation}
\begin{equation}\label{Hi}
  \textbf{H}_{i} = a_{ij}\alpha_{H}\textbf{H}_{j} + b_{ij}\alpha_{H}(\textbf{H}_{j}\cdot\textbf{n}_{ji})\textbf{n}_{ji} + d_{ij}\alpha_{E}(\textbf{n}_{ji}\times\textbf{E}_{j})
\end{equation}

To solve these coupled equations, it is convenient to express the various vector products in Eq.(\ref{Ei}-\ref{Hi}) as matrix products \cite{Bohren,Chaumet}. In detail, by introducing the $3\times3$ matrices $C_{ij}$ and $f_{ij}$, defined as:
\begin{equation}
C_{ij}=
\begin{bmatrix}
a_{ij}+b_{ij}(n_{ij}^x)^2 & b_{ij}n_{ij}^xn_{ij}^y    & b_{ij}n_{ij}^xn_{ij}^z    \\
b_{ij}n_{ij}^yn_{ij}^x    & a_{ij}+b_{ij}(n_{ij}^y)^2 & b_{ij}n_{ij}^yn_{ij}^z    \\
b_{ij}n_{ij}^zn_{ij}^x    & b_{ij}n_{ij}^zn_{ij}^y    & a_{ij}+b_{ij}(n_{ij}^z)^2
\end{bmatrix}
\label{C}
\end{equation}
\begin{equation}\label{F}
f_{ij}=
\begin{bmatrix}
0               & -d_{ij}n_{ij}^z &  d_{ij}n_{ij}^y \\
 d_{ij}n_{ij}^z &  0              & -d_{ij}n_{ij}^x \\
-d_{ij}n_{ij}^y &  d_{ij}n_{ij}^x &  0
\end{bmatrix}
\end{equation}
where $n_{ij}^\beta=\beta_{i}-\beta_{j}$  ($\beta=x,y,$ and $z$) are the components of the direction vector from the $jth$ to the $ith$ particle,  we can re-write Eq.(\ref{Ei}) and Eq.(\ref{Hi}) in the compact form:
\begin{equation}\label{GreenDefinition}
\begin{bmatrix}
\bm{E}_i \\
\bm{H}_i
\end{bmatrix}
=
\begin{bmatrix} 
C_{ij}	&	-f_{ij}\\
f_{ij}	&	C_{ij}\\
\end{bmatrix}
\begin{bmatrix} 
\tilde{\alpha}_E	&	0\\
0	&	\tilde{\alpha}_H\\
\end{bmatrix}
\begin{bmatrix}
\bm{E}_j \\
\bm{H}_j 
\end{bmatrix}
\end{equation}
where $\tilde{\alpha}_E$ and $\tilde{\alpha}_H$ are 3$\times$3 diagonal matrices containing the polarizability $\alpha_E$ and $\alpha_H$ defined by Eq.(\ref{polarizability}) in the case of isotropic materials.

Eq.(\ref{GreenDefinition}) defines the dyadic Green's matrix $\overleftrightarrow{G}_{ij}$ that connects the electromagnetic field of the $ith$-particle with the electromagnetic field of the $jth$-particle. Specifically, $\overleftrightarrow{G}_{ij}$ is obtained as:  
\begin{equation}\label{Gij}
\overleftrightarrow{G}_{ij}= \begin{bmatrix} 
C_{ij}	&	-f_{ij}\\
f_{ij} &	C_{ij}\\
\end{bmatrix}
=
\begin{bmatrix} 
\overleftrightarrow{G}^{ee}_{ij}	&	\overleftrightarrow{G}^{eh}_{ij}\\
\overleftrightarrow{G}^{he}_{ij}  &	\overleftrightarrow{G}^{hh}_{ij}\\
\end{bmatrix}
\end{equation}
The dyadic symbol $\overleftrightarrow{\{\cdots\}}$ is used to stress the fact that we are taking into account all the field components. Therefore, $\overleftrightarrow{G}_{ij}$ is a $6\times6$ matrix. Moreover, one of the advantages of using cgs system unit is that the symmetry relations between electric and magnetic quantities are preserved, i.e. $\overleftrightarrow{G}^{ee}$=$\overleftrightarrow{G}^{hh}$ and $\overleftrightarrow{G}^{eh}$=$-\overleftrightarrow{G}^{he}$ \cite{Chaumet}.
  
The generalization of the formalism for N scatterers is straightforward. Eq.(\ref{Ei}) and Eq.(\ref{Hi}) can be assembled using the a Foldy-Lax scheme such that the local fields at the position of the $ith$ scatterer $\bm{E}^{tot}_i$ and $\bm{H}^{tot}_i$ are the sum of the scattered term of all the other particles plus the incident field ($\textbf{E}_{i,0}$;
$\textbf{H}_{i,0}$) on the $ith$ particle
\begin{equation}\label{FoldyLax1}
\begin{aligned}
\textbf{E}^{tot}_{i} &= \textbf{E}_{i,0} + \sum_{j\neq i}^{N}\alpha_{E,i}C_{ij}\textbf{E}_{j} - \sum_{j\neq i}^{N}\alpha_{H_i}f_{ij}\textbf{H}_{j}=\textbf{E}_{i,0} + \sum_{j\neq i}^{N}\left[\alpha_{E,i}\overleftrightarrow{G}^{ee}_{ij}\textbf{E}_{j} +\alpha_{H_i}\overleftrightarrow{G}^{eh}_{ij}\textbf{H}_{j}\right]
\end{aligned}
\end{equation}
\begin{equation}\label{FoldyLax2}
\begin{aligned}
\textbf{H}^{tot}_{i} = \textbf{H}_{i,0} + \sum_{j\neq i}^{N}\alpha_{H,i}\tilde{C}_{ij}\textbf{H}_{j} +\sum_{j\neq i}^{N}\alpha_{E,i}\tilde{f}_{ij}\textbf{E}_{j}=\textbf{H}_{i,0} + \sum_{j\neq i}^{N}\left[\alpha_{H,i}\overleftrightarrow{G}^{hh}_{ij}\textbf{H}_{j} +\alpha_{E_i}\overleftrightarrow{G}^{he}_{ij}\textbf{E}_{j}\right]
\end{aligned}
\end{equation}
These last two equations can be rewritten as:
\begin{equation}\label{FoldyLaxCompact}
\bm{\Xi}(\bm{r})=\bm{\Xi}_{inc}(\bm{r})+\bm{M}~\bm{\Xi}(\bm{r})
\end{equation}
where $\bm{\Xi}$ is the vector containing the electric $\bm{E}_i$ and magnetic field  $\bm{H}_i$, while $\bm{M}$ is a linear integral operator describing the interactions between the scatterers. To solve Eq.(\ref{FoldyLaxCompact}), successive approximations must be used. The first step is characterized by the Rayleigh-Gans-Debye (RGD) approximation \cite{Bohren,Yurkin}. Within this approximation, $\bm{\Xi}_{inc}(\bm{r})$ is equal to $\bm{\Xi}(\bm{r})$. In this way, we can compute the first-order estimation for every particles. After that, the iterative scheme is obtained by inserting the \emph{jth} interaction of the fields $\bm{\Xi}^{j}(\bm{r})$ into the right side of Eq.(\ref{FoldyLaxCompact}) and evaluating  the next interaction in the left side. The solution of Eq.(\ref{FoldyLaxCompact}) is, therefore,
\begin{equation}
\bm{\Xi}(\bm{r})=\sum_{l=0}^\infty \bm{M}^l~\bm{\Xi}_{inc}(\bm{r})
\end{equation}
which is a direct implementation of the well-know Neumann series 
$$(\bm{I}-\bm{M})^{-1}=\sum_{l=0}^\infty \bm{M}^l$$
where $\bm{I}$ is the unitary operator. A necessary and sufficient condition for the convergence of the Neumann-series is $\left\lVert M \right\rVert<1$. From a physical point of view, this iterative self consistent method lies in successive calculations of interactions between different scatterers. Therefore, the zero order level accounts for no interactions, the first approximation takes into account the influence of the scattering of each dipole on the others once, and so on. 

It is very instructive to write down the compact matrix form of Eq.(\ref{FoldyLax1}) and Eq.(\ref{FoldyLax2}) because it defines the full Green's matrix $\overleftrightarrow{\check{G}}$. Explicitly, the full Green's matrix has the form
\begin{equation}
\overleftrightarrow{\check{G}}= \begin{bmatrix} 
\hat{0}~~&~~\overleftrightarrow{G}_{12}~~&~~\overleftrightarrow{G}_{13}&~~\dots&~~\overleftrightarrow{G}_{1N}\\
\\
\vdots &\ddots& \vdots&\vdots &\vdots\\
\\
\overleftrightarrow{G}_{j1}~~&~~\dots~~&\hat{0}~~&\dots~~&~~\overleftrightarrow{G}_{jN}\\
\\
\vdots &\vdots& \vdots&\ddots &\vdots\\
\\
\overleftrightarrow{G}_{N1}~~&~~\overleftrightarrow{G}_{N2}~~&~~\overleftrightarrow{G}_{N3}&~~\dots&~~\hat{0}\\
\end{bmatrix}
\end{equation}
where $\hat{0}$ represents the $6\times6$ zeros matrix, while the $6\times6$ sub-block are expressed by the matrix (\ref{Gij}).$\overleftrightarrow{\check{G}}$ is a $6N\times6N$ elements where $N$ expresses the total number of scattereres.

Within this formalism, the extinction efficiency of a generic array of scattering particles can be directly obtained from the forward-scattering amplitude using the optical theorem for vector waves for both electric and magnetic polarizations, which results in \cite{Draine}: 
\begin{equation}\label{Crossext}
Q_{ext}=\frac{4\pi k_0}{\pi|E_{inc}|^2NR^2}\sum_{i=1}^N \Im[\bm{p}(\bm{r_i})\cdot\bm{E}^*_{inc}(\bm{r}_i)+\bm{m}(\bm{r}_i)\cdot\bm{H}_{inc}^*(\bm{r}_i)]
\end{equation}
where $\bm{p}(\bm{r}_i)=\alpha_e\bm{E}(\bm{r}_i)$, $\bm{m}(\bm{r}_i)=\alpha_m\bm{H}(\bm{r}_i)$, $R$ is the particle radius, N is the particle number, and the asterisk denotes the complex conjugate. Similarly, the absorption efficiency can be obtained by considering the energy dissipation of both dipoles in the system producing:
\begin{eqnarray}\label{Crossabs}
\begin{aligned}
Q_{abs}=&\frac{4\pi k_0}{\pi|E_{inc}|^2NR^2}\sum_{i=1}^N|\bm{E(\bm{r}_i)}|^2\left(\Im[\alpha_e(\bm{r}_i)]-\frac{2}{3}k_0^3|\alpha_e(\bm{r}_i)|^2\right)\\
&+ \frac{4\pi k_0}{|E_{inc}|^2}\sum_{i=1}^N|\bm{H(\bm{r}_i)}|^2\left(\Im[\alpha_m(\bm{r}_i)]-\frac{2}{3}k_0^3|\alpha_m(\bm{r}_i)|^2\right)
\end{aligned}
\end{eqnarray}
The scattering efficiency can be always obtained by the difference of the extinction and the absorption efficiency, $i.e.$ $Q_{sca}=Q_{ext}-Q_{abs}$. However, this operation requires high numerical accuracy in the computation of both $Q_{ext}$ and $Q_{abs}$ \cite{Chaumet}. To avoid this problem it is possible to directly calculate the scattering efficiency $Q_{scat}$  by evaluating the power radiated in the far field by the oscillating electric and magnetic dipoles, which is \cite{Jackson}:
\begin{equation}\label{Crossscat}
\begin{split}
Q_{sca}&=\frac{k_0^4}{\pi|E_{inc}|^2NR^2}\int\Big(\frac{d\sigma}{d\Omega}\Big)~d\Omega\\&=\frac{k_0^4}{\pi|E_{inc}|^2NR^2}\int\Bigl| \sum_{i=1}^N e^{ik_0\bm{n}\cdot\bm{r}_i}\{\bm{p}(\bm{r}_i)-[\bm{\hat{n}}\cdot\bm{p}(\bm{r}_i)]\bm{\hat{n}}-\bm{\hat{n}\times\bm{m}(\bm{r}_i)}\Bigr|^2~d\Omega
\end{split}
\end{equation}
where $\bm{\hat{n}}$ is an unit vector in the direction of scattering. Moreover, Eq.(\ref{Crossscat}) defines the differential scattering efficiency in the backward and forward direction when $\bm{\hat{n}}$ is equal to the tern $(0,0,-1)$ and $(0,0,1)$, respectively, if the excitation is assumed along the z-axis. Explicitly, the forward and backward scattering efficiencies are defined as:
\begin{equation}\label{Q_fd}
\begin{split}
\left(\frac{d\sigma}{d\Omega}\right)\Bigl|_{\theta=0}&=\left.\frac{4k_0^4}{|E_{inc}|^2NR^2}\frac{d\sigma}{d\Omega}\right\vert_{\theta=0} \\&=\frac{4k_0^4}{|E_{inc}|^2NR^2}\Bigl| \sum_{i=1}^N e^{ik_0\bm{\hat{n}}\cdot\bm{r}_i}\{\bm{p}(\bm{r}_i)-[\bm{\hat{n}}\cdot\bm{p}(\bm{r}_i)]\bm{\hat{n}}-\bm{\hat{n}\times\bm{m}(\bm{r}_i)}\Bigr|^2
\end{split}
\end{equation}
\begin{equation}\label{Q_bs}
\begin{split}
\left(\frac{d\sigma}{d\Omega}\right)\Bigl| _{\theta=\pi}&=\left.\frac{4k_0^4}{|E_{inc}|^2NR^2}\frac{d\sigma}{d\Omega}\right\vert_{\theta=\pi} \\&=\frac{4k_0^4}{|E_{inc}|^2NR^2}\Bigl| \sum_{i=1}^N e^{ik_0\bm{\hat{n}}\cdot\bm{r}_i}\{\bm{p}(\bm{r}_i)-[\bm{\hat{n}}\cdot\bm{p}(\bm{r}_i)]\bm{\hat{n}}-\bm{\hat{n}\times\bm{m}(\bm{r}_i)}\Bigr|^2
\end{split}
\end{equation}
where $\theta$ is the azimuthal angle. 
\section{Scattering properties of elliptic curves and discrete logarithm structures}
\begin{figure}[t!]
\centering
\includegraphics[width=12cm]{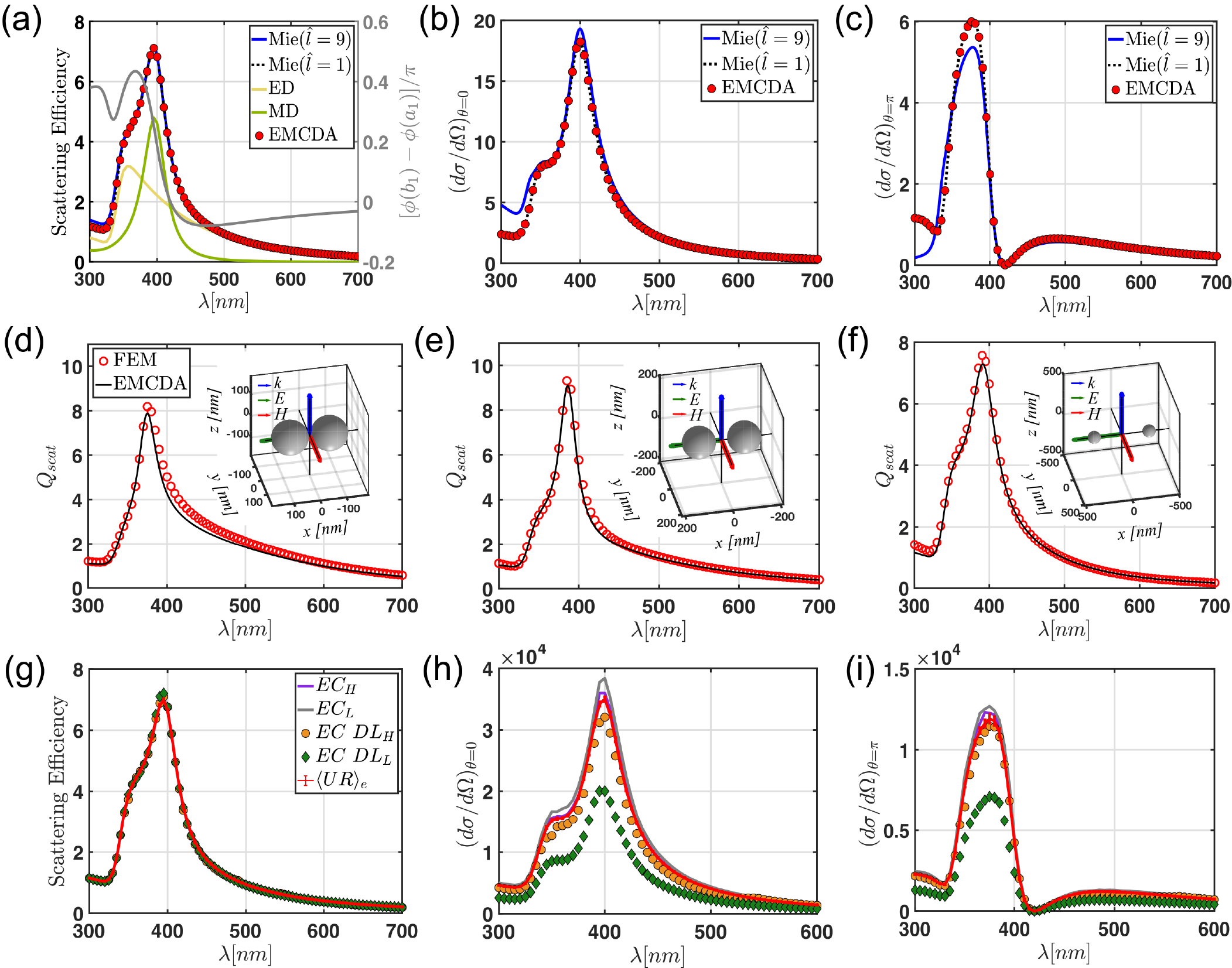}
\caption{(a) Scattering efficiency of a single $TiO_2$ nanoparticle ($R$=70nm) evaluated by using the Mie-Lorentz theory by truncating $\hat{l}$ up to the convergence order provided by  $\hat{l}=x+4.05x^{1/3}+2$ \cite{Wiscombe} ($x$ is the size parameter) as compared to both the analytical result with only the dipolar contribution ($\hat{l}=1$) and the numerical EMCDA calculation (red circle markers). The electric dipole (ED) and the magnetic dipole (MD) contributions are also shown. The grey-left $y$-axis indicates the phase difference $\Delta\phi=\phi(b_1)-\phi(a_1)$, normalized with respect to $\pi$, between the magnetic and electric dipole. Here, $a_1$ and $b_1$ are the Mie-Lorentz coefficients evaluated by using Eq.(\ref{a}) with $\nu=1$. Panels (b-c) show the same benchmark for the differential scattering efficiency in the forward and backward direction, respectively. Panel (d-f) displays the benchmark between the EMCDA and the FEM technique applied to different dimer nanoparticle ($R=$70nm) configurations characterized by an interparticle separation of $10$nm, $50$nm, and $450$nm, respectively. Panels (g-i) show, respectively, the scattering efficiency and the differential scattering efficiency in the forward and backward directions of $EC_H$ (violet line), $EC_L$ (grey line), $EC$ $DL_H$ (circle orange markers), $EC$ $DL_L$ (green diamond markers), and $UR$ (red line) arrays rescaled to avoid touching scatterers. The error bars of the $UR$ case are evaluated as the standard deviation over 20 different disorder realizations.}
\label{Fig13}
\end{figure}
Using the EMCDA framework introduced above we now discuss the scattering properties of aperiodic $TiO_2$ nanoparticles arrays generated according to elliptic curves over $\mathbb{F}_{2111}$ and the corresponding discrete logarithm problem. The magnetic permittivity of the sphere and the surrounding medium is assumed to be 1. All the calculations are performed in air ($n_{host}=1$) under plane wave illumination with $\theta_{inc}$=0$^\circ$ assuming transverse electric polarized light described by
\begin{equation}
\bm{E}=E_0~\hat{\bm{x}}~e^{i\bm{k}\cdot\hat{\bm{z}}}e^{-i\omega~t} ~~~~~~~~~~~~~~~~~~~~~~~\bm{H}=H_0~\hat{\bm{y}}~e^{i\bm{k}\cdot\hat{\bm{z}}}e^{-i\omega~t} 
\end{equation}
where $k=n_{host}2\pi/\lambda$, while the symbol $\{\hat{\cdot}\}$ identifies the unit axes vector.

Before analyzing the scattering properties of the arrays, we performed different benchmarks of the EMCDA method with respect to the analytical full-wave Mie theory \cite{yuyao2018pole,bohren2008absorption} applied to a single $TiO_2$ nanoparticle with a radius of $70nm$. 
Fig.\ref{Fig13} panels (a-c) show the results of this comparison. In particular, panel (a) displays the scattering efficiency computed using the analytical Mie-Lorentz theory by truncating the multipolar expansion  up to the convergence order provided by  $\hat{l}=x+4.05x^{1/3}+2$ \cite{Wiscombe} ($x$ is the size parameter) as compared to both the analytical result with only the dipolar contribution ($\hat{l}=1$) and the numerical EMCDA  calculation (red circle markers). The agreements between the EMCDA and the Mie theory with only the dipolar contribution is almost perfect. Moreover, the relative error due to the dipolar approximation, evaluated from the ratio of area beyond the blue and the black dotted curves, is 1.5\%. Therefore, the scattering properties of these small nanoparticle is sufficiently well-described by considering only the electric ($a_1$) and magnetic ($b_1$) dipole terms of the Mie expansion \cite{Nieto}. Panels (b-c) display, respectively, the differential scattering efficiency in the forward and backward direction evaluated by using Eq.(\ref{Q_fd}) and Eq.(\ref{Q_bs}), respectively, (red circle markers) and compare to the results from the equations:
\begin{equation}\label{MieBF}
Q_{bs}^{Mie}=\frac{1}{(k_0R)^2}\Bigl|\sum_{l=1}^{\hat{l}}(2l+1)(-1)^l(a_l-b_l)\Bigr|^2~~~~~~~~~~~~~~~~~~~~~~~Q_{fd}^{Mie}=\frac{1}{(k_0R)^2}\Bigl|\sum_{l=1}^{\hat{l}}(2l+1)(a_l+b_l)\Bigr|^2
\end{equation}
that are derived from the Mie theory \cite{yuyao2018pole,bohren2008absorption} when $\hat{l}=9$ (blue curve) and $\hat{l}=1$ (dotted black line). Again, the matching between the EMCDA and the Mie theory with $\hat{l}=1$ is excellent. On the other hand, the relative error due to the dipolar approximation is approximately 10\% for both comparisons. This is due to the fact that the higher order multipoles interfere with the dipole moments for a fixed scattering directions. We remark that Eq.(\ref{MieBF}) describes a coherent sum between all the multipole moments. On the other hand, no interference effects contribute to the total scattering efficiency \cite{yuyao2018pole}. Interestingly, Fig.\ref{Fig13} (c) shows that the backscattered light is completely suppressed around $\lambda\sim420nm$, where the relative phase between the electric ($\phi(a_1)$) and magnetic ($\phi(b_1)$) dipoles crosses zero, as shown in the grey $y$-axis of Fig.\ref{Fig13} (a) (see also \cite{Person} for more details). 

In order to analyze the scattering properties of the $EC$ and $EC$ $DL$ arrays, we have selected the structures that showed significantly different modal lifetime behavior. Namely, $EC_H$, $EC_L$, $EC$ $DL_H$, and $EC$ $DL_L$. To avoid the occurrence of overlapping nanoparticles (remember that we are considering now a real scattering object characterize by a size and a material through the polarizabilities expressed by Eq.(\ref{polarizability})), we have rescaled these aperiodic arrays by fixing the minimum particle separation to be of the order of $2R+10nm$. We carefully verified the accuracy of the EMCDA simulations by comparing with simulations performed using the Finite Element Method (FEM) in a dimer  nanoparticles configuration with 10nm gap separation. The FEM model is meshed with 5.6 nm maximum element size and 0.56 nm minimum element size. The total degrees of freedom of the FEM simulation is 1,420,416. We performed the simulations using a 40 core cluster (Intel Xeon(R) CPU E5-2698 v4) with 256Gb total RAM. Typical time to complete full-spectrum simulations was approximately 4 hours and only approximately 3 minutes using the EMCDA method on the same geometry. Moreover, the Green’s matrix spectral method provides fundamental physical information about the light transport properties of open scattering systems that cannot be easily accessed via other numerical methods, such as Finite Difference Time Domain (FDTD) or Finite Elements (FEM). Indeed, in contrast to numerical mesh-based methods, the Green’s matrix spectral method and its EMCDA extension allow one not only to obtain the frequency positions and lifetimes of all the scattering resonances, but also to fully characterize their spectral statistics and measurable scattering parameters. Finally, these methods enable understanding of the full spectral characteristics of the deterministic aperiodic arrays containing several thousand interacting nanoparticles, which are well beyond the reach of mesh-based numerical methods. 

\begin{figure}[h!]
\centering
\includegraphics[width=12cm]{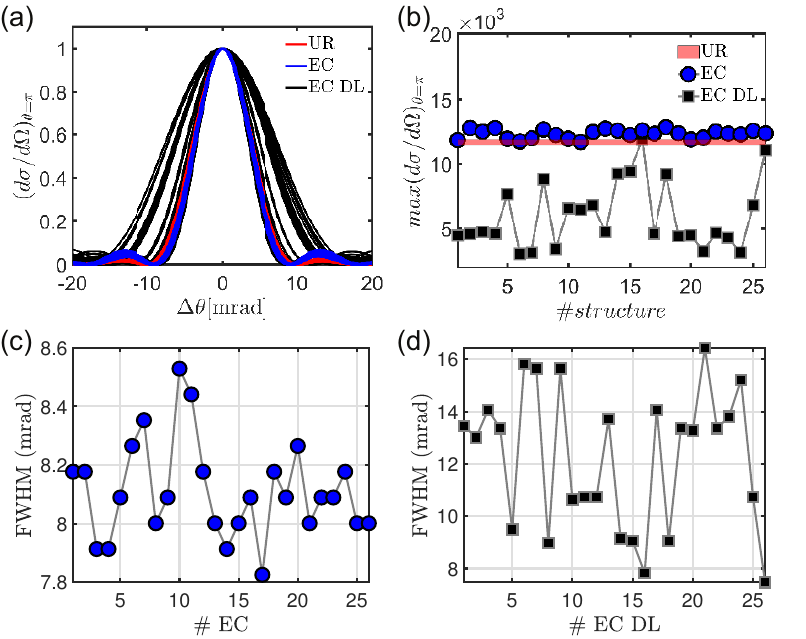}
\caption{ (a) Normalized backscattering cone of different representative $EC$ (blue lines) and $EC$ $DL$ (black lines) structures as compared to 20 different disorder realizations of traditional uniform random arrays (red lines). Table.\ref{Table1} summarizes the averaged structural parameters of the different analyzed devices. Specifically, the $EC$ arrays were selected equidistantly from the 900 different $EC$ structures generated by all the possible combination of the coefficients $A$ and $B$ in the range [1,30] ordered by following the $\hat{\Gamma}$ trend of Fig.\ref{Fig6}. In the same way, the 26 different $EC$ $DL$ arrays were selected equidistantly from the 32 $EC$ $DL$ point patterns, generated as discussed in section \ref{Greensection}, order by following the $\hat{\Gamma}$ trend of Fig.\ref{Fig11}. (b) Intensity peak of the differential scattering efficiency evaluated in the backward direction by using Eq.(\ref{Q_bs}). Panels (c-d) report, respectively, the full width half maximum of the backscattering cone of the $EC$ and $EC$ $DL$ aperiodic arrays.}
\label{Fig14}
\end{figure}

The validation results (shown in Fig.\ref{Fig13} panels (d-f) for the longitudinal polarization) yield a small ($6\%$) discrepancy compared to the ones obtained using our EMCDA method. On the other hand, the results of the EMCDA  analysis on the arrays is reported in Fig.\ref{Fig13} panels (g-i). Since only a small fraction ($<5\%)$ of the particles in the arrays are separated by the $10nm$ minimum gap distance, the application of the EMCDA method to these geometries is fully justified and the contribution of higher-order electromagnetic multipoles can be safely neglected. Our findings show that the scattering spectra of the investigated $EC$ and $EC$ $DL$ structures overlap very well with the spectrum of the ensemble averaged $UR$ system across the entire visible spectrum. This is in agreement with the uncorrelated nature of the $EC$ arrays and of the $EC$ $DL$ arrays with the largest and the smallest modal lifetimes. However, structural differences between these systems can be identified by considering the spectral behavior of their directional scattering parameters. This has been achieved by computing the forward and the backward scattering spectra, which are shown in panels (h) and (i), respectively. In particular, we observe that the data obtained on $EC$ $DL_L$ feature significantly reduced forward and backscattering intensities, reflecting a more correlated spatial structure compared to all the other systems. Smaller differences are also visible among the other $EC$-based structures when compared to the ensemble averaged $UR$ case. 

In order to more precisely address the subtle  modifications in the directional scattering parameters we analyzed in Fig.\ref{Fig14} the linewidth and the maximal differential scattering efficiency in the backward direction for the different arrays. The analysis is performed considering 26 representative $EC$ and $EC$ $DL$ structures. In order to uniformly sample the vast space of structural parameters that we have examined in the section \ref{Greensection}, we have selected 52 aperiodic arrays by using as discriminator the average modal lifetime $\hat{\Gamma}$, as reported in Fig.\ref{Fig6} and Fig.\ref{Fig11}. Specifically, 26 $EC$ arrays were chosen between the 900 different elliptic curves point patterns generated by the integer coefficients $A$ and $B$ in the range [1,30] ordered by following the trend of $\hat{\Gamma}$. Specifically, 26 different $EC$ structures were selected with parameter values that are equidistant between the $EC_L$ and the $EC_H$ structure. In the same way, the 26 $EC DL$ arrays were selected equidistantly between the 36 different structures discussed in Fig.\ref{Fig11}. All these 52 arrays were scaled to avoid the occurrence of overlapping nanoparticles. Table.\ref{Table1} summarizes the averaged structural parameters of all the investigated structures. All share approximately the same minimum and averaged first-neighbor particle separation, as well as the same particle density. 

\begin{table}[h]
\begin{ruledtabular}
\caption{Averaged structural parameters of the different analyzed devices. Specifically, $d_{min}$, $d_1$ and $\rho$ indicates, respectively, the~minimum particle separation, the~averaged first-neighbor particle separation, and~the particle density evaluated as $N/L_xL_y$ ($L_{x,y}$ are the lateral dimension along the $x$ and $y$ direction).}
\centering
\begin{tabular}{cccc}
\toprule
\textbf{Structural Parameters}                         & \boldmath{$EC$}	 & \boldmath{$EC$ $DL$}	& \boldmath{$UR$}\\
\midrule
$\langle d_{min}\rangle$ $[nm]$	 & $152.04\pm1.72$			& $189.79\pm19.35$			& $151.05\pm1.05$\\
$\langle d_{1}\rangle$ $[nm]$	         & $482.99\pm5.03$			& $499.43\pm12.69$			& $484.11\pm3.62$\\
$\rho$    $[\mu m^{-2}]$                     & $1.22\pm0.03$  				& $1.20\pm0.11$ 	        			& $1.23\pm0.04$\\
\bottomrule
\end{tabular} \label{Table1}
\end{ruledtabular}
\end{table}

The normalized lineshapes of the backscattering are displayed in Fig.\ref{Fig14} (a) and show a significant variability. The backscattering of a $UR$ representative realization is also shown for comparison in red. The significant differences in the width of the backscattering angular spectrum (computed at the wavelength of maximum scattering) evidence subtle differences in the structural properties of the arrays that cannot otherwise be resolved by total scattering analysis. The more sensitive interference effects that contribute to the width and intensity of the backscattering cone allow us to differentiate between different $EC$ and $EC$ $DL$ structures for the first time. Note that simply considering the maximum backscattering efficiency, shown in Fig.\ref{Fig14} (b), would not lead to a clear discrimination between a $UR$ and the different $EC$ structures. The full-width-at-half-maximum (FWHM) results obtained for all the investigated structures are plotted in Fig.\ref{Fig14} (c-d), which demonstrate a great variability across the analyzed sample. We should appreciate that the FWHM of the backscatering cone varies by almost a factor of two across the different $EC$ $DL$ structures, which is evidence of significant modifications in the underlying geometrical structure of the arrays. 
Therefore, our findings not only establish that $EC$ and $EC$ $DL$ are remarkably different from uniform random systems, but they may also provide an optical approach to rapidly identify the potential vulnerabilities of modern $EC$-based cryptosystems by investigating coherent light scattering effects in the associated photonic structures.

\section{Conclusions}

In this paper we introduce a novel class of deterministic aperiodic photonic systems that physically implement the distinctive aperiodic order of elliptic curves and their associated discrete logarithm problem. In particular, we addressed structure-property relationships in a large number (900) of aperiodic photonic systems that manifest an extremely rich spectrum of scattering and localization properties that can be engineered to outperform the performances of traditional uniform random media in terms of optical confinement and directional light scattering. By combining the interdisciplinary methods of point patterns spatial statistics with the rigorous Green's matrix solution of the multiple wave scattering problem for electric and magnetic dipoles we systematically explored the spectral and light scattering properties of novel deterministic aperiodic structures with enhanced light-matter coupling for nanophotonics and metamaterials applications to imaging and spectroscopy. By demonstrating significant deviations from traditional random media, our findings not only underline the importance of structural correlations in elliptic curve-based structures for photonics technology but may additionally provide an optics-driven approach to rapidly identify potential vulnerabilities in modern EC-based cryptosystems.

\section*{author contributions}{L.D.N. conceived, supervised and organized the research activities. L.D.N. wrote the manuscript with input from all the authors. Y.C.  performed the numerical simulations, and analyzed the data. F.S. contributed to the development of the simulations methods, performed numerical calculations, and organized the results. All authors contributed to discussions and manuscript revision.}
\section*{funding}{This research was sponsored by the Army Research Laboratory and was accomplished under Cooperative Agreement Number W911NF-12-2-0023. The views and conclusions contained in this document are those of the authors and should not be interpreted as representing the official policies, either expressed or implied, of the Army Research Laboratory or the U.S. Government. The U.S. Government is authorized to reproduce and distribute reprints for Government purposes notwithstanding any copyright notation herein.}
\section*{acknowledgments}{The authors would like to acknowledge Dr. F. Pintore at the Mathematics Department, Oxford University (UK) for fruitful discussions on elliptic curve cryptography. }
%

\end{document}